\documentclass[12pt]{article}

\usepackage{amsfonts}
\usepackage{amsmath}
\usepackage{amssymb}
\usepackage[dvips]{graphicx}
\usepackage{comment}
\usepackage{mathptmx}
\usepackage{color}                    
\usepackage{hyperref}                 
\usepackage[normalem]{ulem}

\newcommand{\ba}{\begin{eqnarray}}
\newcommand{\ea}{\end{eqnarray}}
\newcommand{\nn}{\nonumber}

\newcommand{\cT}{\mathcal{T}}
\newcommand{\hC}{{\hat C}}
\newcommand{\I}{\mathrm{i}}
\newcommand{\lb}{\left<\!\!\left<}
\newcommand{\lbl}{\left<\hskip-0.36em\left<\hskip0.1em}
\newcommand{\lbg}{\left<\hskip-0.46em\left<\hskip0.1em}
\newcommand{\rb}{\right>\!\!\right>}
\newcommand{\rbl}{\right>\hskip-0.35em\right>\hskip0.1em}
\newcommand{\rbg}{\right>\hskip-0.45em\right>\hskip0.1em}
\newcommand{\cX}{\mathcal{X}}

\newcommand{\hide}[1]{}
\newcommand{\pj}{{\cal P}}

\begin{document}

\begin{titlepage}

\begin{center}


\vfill

\textbf{\LARGE
Dimensional oxidization on coset space}

\vfill
{\large
Koichi Harada$^\dagger$\footnote{
e-mail address: harada@hep-th.phys.s.u-tokyo.ac.jp},
Pei-Ming Ho$^\ddagger$\footnote{
e-mail address: pmho@phys.ntu.edu.tw}, 
Yutaka Matsuo$^{\diamondsuit}$\footnote{
e-mail address:
matsuo@phys.s.u-tokyo.ac.jp} and
Akimi Watanabe$^\dagger$\footnote{
e-mail address: awatanabe@hep-th.phys.s.u-tokyo.ac.jp}
}
\vskip 10mm
{\it
$^\dagger$
Department of Physics, Faculty of Science, University of Tokyo,\\
Hongo 7-3-1, Bunkyo-ku, Tokyo 113-0033, Japan
}\vskip 1mm
{\it
$^\ddagger$
Department of Physics and Center for Theoretical Physics, \\
National Taiwan University, Taipei 106, Taiwan,
R.O.C.}\vskip 1mm
{\it $^\diamondsuit$
Department of Physics \& Trans-scale Quantum Science Institute\\ \& Mathematics and Informatics Center, University of Tokyo,\\
Hongo 7-3-1, Bunkyo-ku, Tokyo 113-0033, Japan}

\vfill

\begin{abstract}
In the matrix model approaches of string/M theories, 
one starts from a generic symmetry $gl(\infty)$ to reproduce the space-time manifold.
In this paper, we consider the generalization in which the space-time manifold emerges from a gauge symmetry algebra which is not necessarily $gl(\infty)$.
We focus on the second nontrivial example after the toroidal compactification, the coset space $G/H$, and propose a specific infinite-dimensional symmetry which realizes the geometry. It consists of the gauge-algebra valued functions on the coset and Lorentzian generator pairs associated with the isometry. 
We show that
the $0$-dimensional gauge theory 
with the mass and Chern-Simons terms gives the gauge theory on the coset with scalar fields associated with $H$.
\end{abstract}
\vfill

\end{center}

\end{titlepage}

\setcounter{footnote}{0}

\section{Introduction}

A description of the space-time by matrices has been an essential subject in gauge/string theories.  The origin of the idea may go back to the large-$N$ reduced model \cite{Eguchi:1982nm} for the gauge theories. In string theory, a description of the dimension by the infinite number of D-branes was proposed in Ref.\cite{Taylor:1996ik}. Since the late 90s, the matrix model began to be an alternative description of M-theory \cite{Banks:1996nn},  and string theory \cite{IKKT}. While in the beginning, they describe only the flat $d$-dimensional torus, later the management of the coset or homogeneous spaces were explored (for instance, in Refs.\cite{Kitazawa:2002xj, Kawai:2010sf}).

In this paper, we explore a more direct correspondence between the gauge symmetry algebra and the space-time manifold by relaxing the requirement that the gauge symmetry algebra be identical to the matrix algebra.
The motivation comes from our study of the higher-dimensional Yang-Mills theory from an infinite-dimensional algebra \cite{Ho:2009nk}. A prototype example is the $S^1$ compactification where the current (Kac-Moody) algebra described by the gauge algebra functions on $S^1$, $\mathcal{T}^A_n$, where the integer $n$ is a label for the KK-modes and $A$ for the gauge algebra $\mathfrak{k}$. We also need to include extra generators $u$ and $v$, where $v$ is the central charge, and $u$ is the level operator. The algebra takes the form
\begin{equation}
\left[\mathcal{T}^A_{n}, \mathcal{T}^B_m\right]= \I {F^{AB}}_C \mathcal{T}^C_{n+m} +nv\,G^{AB} ,\quad
\left[u, \mathcal{T}^A_n\right]=n\mathcal{T}^A_n\,.
\end{equation}
We note that the additional generator $u$ describes the derivative with respect to $\theta$. The center $v$ is necessary to generate the KK-mass. As a straightforward generalization, one may obtain an algebraic description of  the toroidal compactification $T^n$ from the $n$-loop algebra\footnote{The idea of using a Lorentzian pair $u,v$ came from 
the BLG-type
formulation of M2-branes (for instance, see Ref.\cite{Ho:2008ei}) as well as M5-branes \cite{Ho:2008nn,Ho:2008ve}. In the traditional approach to the matrix model, the generator $v$ is ignored, and the associated equation of motion is absent.}

In this paper, as the next simplest example, we consider the algebra associated with the coset space. The algebra consists of $\mathfrak{k}$ (the gauge algebra)-valued function on $G/H$. As in the Kac-Moody case, we need the extra generators, which describe the isometry $G$ and the analogue of the center. For simplicity, we restrict ourselves to the $0$-dimensional Yang-Mills theory with an additional mass term and Chern-Simons-like coupling (which is not restricted to 3 dimensions), and show that it produces a Lagrangian of the gauge fields coupled with adjoint scalar fields on $G/H$. 

We organize the paper as follows. In section 2, we give a brief review of the differential geometry of the coset space, along the line of Ref.\cite{Castellani:1999fz}. We use the embedding function of $G/H$ into $G$. The arbitrariness of the embedding suggests that the group $H$ behaves as an extra gauge symmetry.
In section 3, we propose the infinite-dimensional algebra associated with the coset space $G/H$ and the gauge symmetry $K$. In the definition of the algebra, there is no ambiguity associated with the embedding. In sections 4 and 5, we derive IKKT-type Lagrangian with mass and Chern-Simons terms with the infinite-dimensional symmetry. With a Higgs-like mechanism, we obtain a field theory Lagrangian defined on the coset. While it has the $K$ gauge symmetry manifestly, the $H$ gauge symmetry is hidden. In sections 6, 7, 8, we decompose the original field variables $X$ into the gauge and the scalar fields on the coset, by appropriately choosing the basis of the tangent space of $G$ into the tangent (gauge fields) on the coset $G/H$ and orthogonal (scalar fields) directions. At this level, the Lagrangian has both the $K$ and $H$ gauge symmetries.\footnote{We note that the gauge theory on a coset space $G/H$ was constructed as a matrix model in Ref.\cite{Kawai:2010sf}.
They start from the gauge theory on the group manifold $G$, and obtained the action on $G/H$ as a dimensional reduction. On the other hand, we start from the definition of the infinite-dimensional algebra associated with the coset and derived directly the coset action.}

\section{A brief review of the cosets $G/H$}

In this section,
we review basics about the geometry of coset spaces $G/H$ along the line of Ref.\cite{Castellani:1999fz}.

\subsection{Lie algebra decomposition}
A coset is a quotient space $G/H$
of a Lie group $G$ over a subgroup $H \subset G$.
Let $\mathfrak{g}$ and $\mathfrak{h}$ be the Lie algebras of $G$ and $H$,
respectively,
we can then decompose the Lie algebra $\mathfrak{g}$ of $G$ as
\ba
\mathfrak{g}=\mathfrak{h}\oplus \mathfrak{m}
\ea
where $\mathfrak{m}$ represents the coset part of the Lie algebra.

When both $G$ and $H$ are compact Lie groups,
one may choose $\mathfrak{m}$ such that
\ba
\left[\mathfrak{h}, \mathfrak{m}\right] \subseteq \mathfrak{m},
\label{hmm}
\ea
with $\mathfrak{m}$ and $\mathfrak{h}$ orthogonal to each other
with respect to the inner product.
This means that one can choose the generators
$T_a$ ($a, b=1,\cdots, |G|$) of $\mathfrak{g}$
such that 
$T_i\in \mathfrak{h}$ for $i = 1, 2, \cdots, |H|$
and $T_\alpha\in \mathfrak{m}$ for $\alpha = |H|+1, \cdots, |G|$,
and that the Lie bracket of $\mathfrak{g}$
\begin{equation}
\left[ T_a, T_b\right] =\I {f_{ab}}^c T_c \, 
\end{equation}
is decomposed as
\ba
\left[T_i, T_j\right]&=&
\I{f_{ij}}^k T_k \, ,
\\
\left[T_i, T_\alpha\right]&=&
\I{f_{i\alpha}}^\beta T_\beta \, ,
\\
\left[T_\alpha, T_\beta\right]&=&
\I{f_{\alpha\beta}}^i T_i+\I{f_{\alpha\beta}}^\gamma T_\gamma \, .
\ea
In other words,
the structure constants ${f_{ij}}^\alpha$, ${f_{i\alpha}}^j$ vanish.
In the following,
we will not require $G$ and $H$ to be compact Lie groups,
but we will assume that eq.\eqref{hmm} holds.

\subsection{Coset representative}
\label{Coset-rep}

An element $x \in G/H$ can be identified with $x=gH$ 
for some $g\in G$.
It is natural to define a projection $\pi: G\rightarrow G/H$ as
\ba
\pi (g)= gH \, ,
\ea
and an embedding map
$\sigma: G/H \rightarrow G$ such that
$\pi \cdot \sigma: G/H \rightarrow G/H$ is the identity map.
$\sigma(x)$ can be viewed as the representative of $x \in G/H$.
But the choice of the coset representative is not unique: 
one may change it as
\begin{equation}\label{h-gauge}
\sigma(x) \quad \rightarrow \quad
\tilde{\sigma}(x)=\sigma(x)\tilde{h}(x)
\end{equation}
for an arbitrary map $\tilde{h}: G/H \rightarrow H$.
In the following,
we will refer to this arbitrariness of $\sigma$ as the ``$H$-gauge symmetry".

The multiplication by $g \in G$ to $x \in G/H$ from the left
is an isometry transformation on the coset $G/H$.
For any $g \in G$,
there is a function $h(g, x)$ on $G/H$ such that
\begin{equation}\label{h-twist}
g \sigma(x) = \sigma(gx) h(g, x),\qquad h(g, x)\in H \, .
\end{equation}
The $H$-twist $h(g,x)$ satisfies the cocycle condition
\ba
h(g_1 g_2, x)=h(g_1, g_2 x)h(g_2, x)
\ea
as a consequence of the associativity of $G$.

\subsection{Adjoint representation}

In the following, 
we will use the notation $D(g)$ for the adjoint representation
so that
\ba
\label{adjoint}
g^{-1}T_a g={D_a}^b(g)T_b \, .
\ea
As a representation,
it satisfies ${D_a}^b(g_1 g_2)={D_a}^c(g_1) {D_c}^b(g_2)$.

Assuming that $\mathfrak{g}$ is a 
Lie algebra equipped with a non-degenerate invariant inner product
$\langle T_a, T_b \rangle$.
Without loss of generality, 
we assume that
$\langle T_a, T_b \rangle = \eta_{ab}$,
where $\eta_{ab}$ is a diagonal matrix with eigenvalues $\pm 1$.
\footnote{
For a semi-simple Lie-algebra,
one can take the inner product to be the Killing form.
For an Abelian group such as $U(1)^n$,
one can take any $\eta_{ab}$.
}
The invariance of the inner product implies
the orthogonality of the adjoint representation:
\ba
\eta_{ab} \ {D_c}^a(g){D_d}^b(g)=\eta_{cd} \, .
\ea
The transpose
\begin{equation}
{\bar{D}_a}^{ \,\  b}(g) \equiv \eta^{bc}\eta_{ad} {D_c}^d(g)
\end{equation}
of $D(g)$ is also its inverse, i.e.
\begin{equation}
{D_a}^c(g) {\bar{D}_c}^{\ \,b}(g) ={\bar{D}_a}^{\ \,c}(g) {D_c}^{b}(g)=\delta_a^b\, .
\end{equation}

\subsection{Vielbein and $H$-connection}

Given a local coordinate system $\{\theta^\mu\}$
($\mu = 1, 2, \cdots |G| - |H|$) of $G/H$,
we shall denote $\sigma(x(\theta))$ simply as $\sigma(\theta)$ on the local patch.
Define the covariant frame (vielbein) $V_\mu{}^\alpha(\theta)$
and the $H$-connection $\Omega_\mu{}^i(\theta)$ by
\begin{equation}\label{E-def}
E_\mu\equiv\sigma(\theta)^{-1}\partial_\mu \sigma(\theta)=
\I V_\mu{}^\alpha (\theta)T_\alpha +\I \Omega_\mu{}^i(\theta) T_i \, ,
\end{equation}
which can be expressed as an equation of 1-forms:
$E=\sigma^{-1}d\sigma=\I(V^\alpha T_\alpha+\Omega^i T_i)$.
The inverse vielbein $V_{\alpha}{}^{\mu}$ by definition satisfies
\ba
V_\mu{}^\alpha(\theta)V_\alpha{}^\nu(\theta)=\delta_\mu^\nu \, ,
\qquad
V_\alpha{}^\mu(\theta)V_\mu{}^\beta(\theta)=\delta^\beta_\alpha \, .
\ea

The invariant inner product of $\mathfrak{g}$ induces a metric on $G/H$:
\ba
g_{\mu\nu}(\theta)\equiv
\eta_{\alpha\beta}V_\mu{}^\alpha(\theta) V_\nu{}^\beta(\theta)\,,
\quad
g^{\mu\nu}(\theta)\equiv
\eta^{\alpha\beta}V_\alpha{}^\mu(\theta) V_\beta{}^\nu(\theta)\,,
\label{metric-g}
\ea
where $\eta^{\alpha\beta}$ is the inverse of $\eta_{\alpha\beta}$
and $g^{\mu\nu}$ is the inverse of $g_{\mu\nu}$.

Under the $H$-gauge transformation (\ref{h-gauge}),
$E$ transforms as
$E \rightarrow 
\tilde E \equiv \tilde{h}^{-1}E\tilde{h} + \tilde{h}^{-1}d\tilde{h}$,
which implies
\begin{align}
V_\mu(\theta) &\rightarrow 
\tilde{V}_\mu(\theta) \equiv \tilde{h}^{-1}(\theta)V_\mu(\theta) \tilde{h}(\theta) \, ,
\label{h-gauge2-V}
\\
\Omega_\mu(\theta) &\rightarrow
\tilde{\Omega}_{\mu}(\theta) \equiv
\tilde{h}^{-1}(\theta) \Omega_\mu(\theta) \tilde{h}(\theta)
- \I \tilde{h}^{-1}(\theta)\partial_\mu \tilde{h}(\theta) \, .
\label{h-gauge2-O}
\end{align}

The isometry transformation (\ref{h-twist}) may be written as
\ba
\sigma(\theta^g)=g\sigma(\theta)h^{-1}(\theta,g)
\label{h-twist-1}
\ea
where $\theta^g$ is the coordinate of $gx(\theta)$.
It leads to a transformation of $E$ as 
\ba
\sigma(\theta^g)^{-1}d\sigma(\theta^g)=
h(\theta,g) (\sigma^{-1}(\theta)d\sigma(\theta))h^{-1}(\theta,g)+h(\theta, g)dh^{-1}(\theta, g).
\ea 
Picking up the $T_\alpha$ components, we find
\ba
V^\alpha(\theta^g)=V^\beta(\theta) {D_\beta}^\alpha(h^{-1}(\theta,g)) \, .
\ea
The metric $g_{\mu\nu}d\theta^\mu d\theta^\nu$ is thus 
manifestly invariant under the left $g$-action.


The Maurer-Cartan equation for $E$ is
\ba
dE + E\wedge E=0 \, ,
\ea
which decomposes into $\mathfrak{m}$ and $\mathfrak{h}$ as
\ba
dV^\alpha + \frac\I2 {f_{\beta\gamma}}^\alpha V^\beta\wedge V^\gamma
+ \I{{f_{i\beta}}^\alpha}\Omega^i\wedge V^\beta &=&0 \, ,
\\
d\Omega^i +\frac{\I}{2}{f_{jk}}^i \Omega^j\wedge \Omega^k
+ \frac\I2 {f_{\alpha\beta}}^i V^\alpha\wedge V^\beta &=&0 \, .
\label{Omega-field}
\ea
The last equation says that the field strength
of the $H$-connection $\Omega$ is non-zero
whenever $f_{\alpha\beta}{}^i \neq 0$.

\subsection{Infinitesimal isometry transformation}

We consider the infinitesimal version of
the isometry transformation \eqref{h-twist} on the coset $G/H$
with 
\begin{equation}
g = 1+\epsilon^a T_a \, .
\end{equation}
Eq.(\ref{h-twist-1}) implies that
\ba
h(\theta, g)&=& 1+\epsilon^a {\Lambda_a}{}^i(\theta) T_i \, ,
\\
(\theta^g){}^\mu &=&
\theta^{\mu} - \epsilon^a {u_a}^\mu(\theta) \, 
\ea
for some functions $\Lambda_a{}^i(\theta)$ and $u_a{}^\mu(\theta)$.
The infinitesimal form of (\ref{h-twist}) becomes
\ba\label{infiso}
T_a \sigma(\theta)= 
-\hat u_a(\theta) \sigma(\theta)+\sigma(\theta){\Lambda_a}^i(\theta) T_i \, ,
\ea
where 
\begin{equation}\label{uhat-def}
\hat u_a(\theta)={u_a}^\mu(\theta)\partial_\mu \, .
\end{equation}
The first term on the right hand side describes
an infinitesimal variation on the coset space $G/H$.

Multiplying $\sigma^{-1}$ from the left on both sides of eq.(\ref{infiso}),
one obtains
\ba
\sigma^{-1}T_a\sigma &=&
{D_a}^b(\sigma)T_b =
 {D_a}^\alpha(\sigma)T_\alpha +{D_a}^i(\sigma)T_i
\ea
on the left-hand side according to eq.(\ref{adjoint}),
and deduces from eq.(\ref{E-def}) that
\ba
{D_a}^\alpha(\sigma(\theta))&=&
-\I{u_a}^\mu(\theta){V_\mu}^\alpha(\theta) \, ,
\\
{D_a}^i(\sigma(\theta))&=&
-\I{u_a}^\mu(\theta) {\Omega_\mu}^i(\theta)+{\Lambda_a}^i(\theta) \, .
\ea
Thus we can solve $u_a{}^{\mu}$ and $\Lambda_a{}^i$ as
\ba
{u_a}^\mu(\theta) &=&
\I\, {D_a}^\alpha(\sigma(\theta)) {V_\alpha}^\mu(\theta) \, ,
\\
{\Lambda_a}^i(\theta)&=& 
{D_a}^i(\sigma(\theta))+\I \,{u_a}^\mu(\theta){\Omega_\mu}^i(\theta) \, .
\ea

As eq.(\ref{infiso}) tells us how $T_a$ acts on $\sigma$,
this action must realize the Lie algerba
$\left[T_a, T_b\right]=\I {f_{ab}}^c T_c$,
which then implies that
\begin{align}
& {u_a}^\mu\partial_\mu {u_b}^\nu-{u_b}^\mu\partial_\mu {u_a}^\nu
=\I {f_{ab}}^c {u_c}^\nu,\label{upu}
\\
& {u_a}^\mu\partial_\mu \Lambda_b
-{u_b}^\mu\partial_\mu \Lambda_a+\left[\Lambda_a,\Lambda_b\right]
=\I {f_{ab}}^c \Lambda_c \, ,
\label{upL}
\end{align}
where $\Lambda_a \equiv \Lambda_a{}^i T_i$.

A measure $d\Theta$ can be defined on the coset space $G/H$
such that it is invariant under isometry transformations.
For an arbitrary 
normalizable
regular function $f(\theta)$ on $G/H$,
we have
\ba
\int d\Theta \ {u_a}^\mu(\theta) \partial_\mu f(\theta)=0 \, .
\ea
For later use,
we introduce an orthonormal basis
$\{\lambda_\Xi(\theta)\}$ on $G/H$ for which
\ba
\int d\Theta\  \lambda_\Xi(\theta) \lambda_\Pi(\theta) = \delta_{\Xi\Pi} \, .
\label{ortho-basis}
\ea
When the group $G$ is compact, we have a discrete family of the basis. For the noncompact case, we have to reinterpret the labels $\Sigma, \Pi$ to be continuous. While it is easier to restrict $G$ to be compact, we do not see a strong obstacle for the generalization to the noncompact cases.

\section{An infinite dimensional Lie algebra for $G/H$}

In this section,
we introduce a new class of infinite-dimensional Lie algebras
that is a generalization of the infinite-dimensional Lie algebras
considered in Ref.\cite{Ho:2009nk}
that were used to promote the base space of
a Yang-Mills theory to higher dimensions.

The infinite-dimensional Lie algebra is associated with 
a Lie group $K$ and a coset space $G/H$ as follows.
We will use the notation for $G$ and $H$ as in the previous section.
The Lie algebra $K$ is defined in terms of the basis $\mathbf{T}^A$ as
\begin{equation}
\left[ \mathbf{T}^A, \mathbf{T}^B\right]= \I {F^{AB}}_C \mathbf{T}^C,
\end{equation}
with the invariant metric $G^{AB}=\left(\mathbf{T}^A, \mathbf{T}^B\right)$
with $A,B=1,\cdots |K|$. We denote the Lie algebra associated with $K$ as $\mathfrak{k}$. 
The group $K$ and the coset $G/H$ can be chosen independently.

We will refer the algebra for the coset space as $\hat C(K; G, H)$,
or simply $\hC$. 
The generators of $\hat C(K; G, H)$ consist of $u_a, v^a$ ($a=1,\cdots, |G|$)
associated with the isometry,
and the infinite number of generators $\cT^A[\lambda]$ 
($A = 1, 2, \cdots, |K|$), which depends linearly on any regular function $\lambda(\theta)$ on $G/H$.

We define the Lie bracket by the following relations:
\ba
\label{eq:extension}
\left[ \cT^A[\lambda_1], \cT^B[\lambda_2]\right]&=&
\mathrm{i} {F^{AB}}_C \cT^C[\lambda_1\lambda_2] 
 -\sum_{a=1}^{|G|} G^{AB}v_a \int d\Theta \ 
 \lambda_1(\theta) (\hat u_a \lambda_2(\theta)) \, ,\label{FFcom}
\\
\label{uTcom}
\left[u_a , \cT^B[\lambda]\right]&=&
\cT^B[\hat u_a \lambda] \, ,
\\
\left[u_a, u_b\right] &=&
\mathrm{i} {f_{ab}}^c u_c \, ,
\\
\label{uvcom}
\left[u_a, v^b\right] &=& -\mathrm{i} {f_{ac}}^b v^c \, ,
\\
\left[v^a, v^b\right]&=& \left[v^a, \cT^A[\lambda]\right]=0 \, ,\label{vvcom}
\ea

We have two structure constants: ${F^{AB}}_C$ for the Lie algebra $\mathfrak{k}$ (for the gauge symmetry), and ${f_{ab}}^c$ for $\mathfrak{g}$ (for the isometry of the coset). 
We may roughly identify $\mathcal{F}^A[\lambda]$ with a $\mathfrak{k}$-valued function on $G/H$, say $\mathbf{T}^A\lambda(\theta)$ where $\mathbf{T}^A\in \mathfrak{k}$.
The first term in eq.(\ref{FFcom}) is the algebra for such functions. 
The relation (\ref{uTcom}) implies that $u_a$ acts as an infinitesimal $G$ isometry on the functions on $G/H$, written in the form of $\hat u_a=u_a^\mu\partial_\mu$, which acts on the function $\lambda$. The last line (\ref{vvcom}) and the second term of eq.(\ref{FFcom}) show that $v^a$ may be regarded as an analogue of the central extension, but it has a nonvanishing commutator with $u_a$ when $G$ is non-Abelian as in eq.(\ref{uvcom}).
The dependence of the algebra on $H$ comes implicitly from the facts that $\lambda$ is a function on $G/H$, and $\hat u_a$ is realized as a differential operator acting on it.

The Lie bracket defined above satisfies the Jacobi identity
\begin{align}
[A, [B, C]] + [B, [C, A]] + [C, [A, B]] = 0
\end{align}
for any three generators $A, B, C$ in $\hat{C}(K; G, H)$.

The invariant inner product on $\hat C(K; G, H)$ is defined by
\ba
\label{eq:Killing1}
\left<\cT^A[\lambda_1], \cT^B[\lambda_2]\right>&=& G^{AB} \int d\Theta\, \lambda_1(\theta) \lambda_2(\theta) \, ,
\\
\label{eq:Killing2}
\left< u_a, v^b\right> &=&  \delta^b_a \, ,
\label{eq:LorentzIP}
\ea
with other combinations vanishing.
For any three generators $A, B, C$ in $\hat{C}(K; G, H)$,
the invariant inner product satisfies
\begin{align}
\left< [A, B], C \right> + \left< B, [A, C] \right> = 0 \, .
\end{align}
The Jacobi identity and the invariance of the inner product
are examined in appendix \ref{a:Jacobi}.
We note that this invariant inner product is not positive-definite
because of eq.(\ref{eq:LorentzIP}).
An inner product with the Lorentzian-signature
was considered previously in Ref.\cite{Ho:2009nk}, but not very common in the literature of the matrix models.
We will see below that a Yang-Mills theory dimensionally reduced to 0 dimension with
the algebra $\hat{C}(K; G, H)$
is equivalent to a gauge field theory living on the coset $G/H$
as the base space.

We note that the algebra does not have an ambiguity
in the choice of the coset representative $\sigma(\theta)$. 
The isometry generator $\hat u_a$
is invariant under the $H$-gauge transformation, 
as we will show later (\ref{Cbasis-1}, \ref{h-gauge3})
in terms of the coordinates $\theta$.

We remark that our approach based on $\hat C$ is slightly different from the conventional form of the matrix model. For instance, in the case of the noncommutative $\mathbb{T}^2$ (see, for instance a review article \cite{Konechny:2000dp}), one defines the matrix algebra by two generators $U_1$, $U_2$, satisfying $U_1 U_2= e^{2\pi i\theta} U_2 U_1$. The matrix algebra $T_\theta$ is defined by its envelopping algebra generated by $(U_1)^{n_1}(U_2)^{n_2}$, ($n_1, n_2\in \mathbb{Z}$). In order to reproduce the Yang-Mills theory on $\mathbb{T}^2$, we have to add, extra generators $X_1, X_2$ satisfying $[X_i, U_j]=-2\pi \I \delta_{ij} U_j$, which act as endomorphism on $T_\theta$. In our case, the analogue of $T_\theta$ is generated by $\mathcal{T}^{A}[\lambda]$ which is the basis of the functions on the coset. The extra endomorphism generators $X_i$ in $T_\theta$ are denoted by the elements $u_a$. At the same time, we include $v^b$ which is not included in the matrix model, but are necessary to obtain the proper equation of motion for the non-Abelian coset space. In all, our treatment gives a generalization of the matrix model, which is applicable to the general coset space
without imposing constraints on the field variables.

\subsection{Examples}

\begin{itemize}

\item
When $G=U(1)$ with $H$ being the trivial group
composed of nothing but the identity element $id$,
the algebra $\hat C(K; U(1), id)$ becomes
\ba
\left[\cT^A_n, \cT^B_m\right]&=&
\mathrm{i} {F^{AB}}_C \cT^C_{n+m}+G^{AB} n v \delta_{n+m} \, , \\
\left[ u, \cT^A_n\right]&=& n \cT^A_n \, ,
\ea
where we identify $\cT^A_n=\mathbf{T}^A e^{\I n\theta}$,
$\hat u=-\I\partial_\theta$ 
and $\int d\Theta=\frac{1}{2\pi}\int_0^{2\pi} d\theta$.
In this case,
$\hat C(K; U(1), id)$ is an affine Lie algebra $\hat{\mathfrak{k}}$,
where $v$ is the center and $u$ is the level operator.
It was shown in Ref.\cite{Ho:2009nk} that,
a Yang-Mills theory on $D$-dimensional base space
with the gauge symmetry algebra $\hat C(K; U(1), id)$
is equivalent to a Yang-Mills theory on $(D+1)$-dimensional space
with the gauge group $K$.

\item
Similarly, for $G=U(1)^{\otimes \ell}$, the algebra becomes
\ba
\left[\cT^A_{\vec n}, \cT^B_{\vec m}\right]&=& 
\mathrm{i} {F^{AB}}_C \cT^C_{\vec n+\vec m}
+G^{AB}\sum_a n_av_a \delta_{\vec n+\vec m} \ , \\
\left[ u_a, \cT^A_{\vec n}\right]&=& n_a \cT^A_{\vec n} \, .
\ea
This may be referred to as the $\ell$-loop algebra 
which has $\ell$ central extensions $v^a$.
The generators are labeled by $\vec n\in \mathbb{Z}^\ell$. 
It should be clear from Ref.\cite{Ho:2009nk} that
a $D$-dimensional Yang-Mills theory
with the gauge symmetry algebra $\hat C(K, U(1)^{\otimes \ell}, id)$
is equivalent to a $(D+\ell)$-dimensional Yang-Mills theory
with the gauge symmetry group $K$.
Mathematically,
the representation theory for the cases of $\ell=1,2$ are well-known.
For $\ell=2$,
the algebra is called ``toroidal algebras'',
whose q-deformation \cite{r:toroidal} were intensively studied recently
in the context of the AGT conjecture.
For $\ell>2$, 
not much is known from the mathematical study on the representations.

\item
The isometry group $G$ is Abelian in the examples above.
The simplest non-Abelian example is the coset
$G/H = SU(2)/U(1) = S^2$.
See appendix \ref{cosetS2} for details of the coset description.
\footnote{Matrix model description of fuzzy sphere goes back to Ref.\cite{CarowWatamura:1998jn} .}
We will consider the 0-dimensional gauge theory
with the symmetry algebra $\hat C(K; SU(2), U(1))$ in Sec.\ref{sec-S2}.
It will be shown that the Yang-Mills theory reduced to $0$ dimension
with the symmetry algebra $\hat C(K; SU(2), U(1))$,
supplemented with cubic and quadratic terms,
is equivalent to a Yang-Mills theory on $S^2$
with the symmetry group $K$,
including a generalized Chern-Simons term
and a massive scalar field.

\end{itemize}

\section{Gauge theory with symmetry algebra $\hat C$}

We will focus on 
Yang-Mills theories dimensionally reduced to $0$ dimension,
to study the dimensional oxidization by $\hC$.
As we will focus on the bosonic sector,
it can also be called the Yang-Mills matrix model \cite{Steinacker:2019fcb}
or the reduced model \cite{Kawai:2010sf}.
More general analysis of the super Yang-Mills theory
coupled with adjoint scalars was made in Ref.\cite{Ho:2009nk}
for the higher loop algebras.

The bosonic part of the IKKT model coincides with
the Yang-Mills theory dimensionally reduced to 0 dimension.
It has the action
\ba\label{IKKT}
S_0=\frac{1}{4}\sum_{I,J=1}^d \left<[\cX_I, \cX_J], [\cX^I, \cX^J]\right> \, ,
\ea
where the indices $I, J$ are raised and lowered using
a metric $\eta_{IJ}$ which is diagonal with the eigenvalues $\pm 1$.
The variables $\cX^I$ ($I=1,\cdots, d$) are typically infinite-dimensional matrices.
Our general strategy is to replace the algebra $gl(\infty)$ 
of infinite-dimensional matrices by $\hC(K; G, H)$
to obtain the dimensional oxidization on the coset space $G/H$.
Taking values in $\hC(K; G, H)$, we expand
the matrices $\cX_I$ in terms of the generators as
\ba
\label{Xcomp}
\cX_I= X_I + Y_I + Z_I \, ,
\ea
where
\ba
X_{I} = \sum_\Xi  X_{IA\Xi}\cT^A[\lambda_{\Xi}] \, ,
\quad
Y_{I} = \sum_a Y_{Ia} v^a \, ,
\quad
Z_{I} = \sum_a Z_{I}^a u_a \, .
\ea
Here,
$\sum_{\Xi}$ is a sum over an orthonormal basis $\{\lambda_\Xi\}$ on $G/H$.
We will refer to $Y, Z$ as the ``ghosts''.

To derive a more explicit expression for the action (\ref{IKKT}),
we expand the commutator $\left[\cX_I, \cX_J\right]$ as
\ba
\left[\cX_I, \cX_J\right]&=&
X_{IA\Xi_1} X_{JB\Xi_2}\left(
\I {F^{AB}}_C \cT^C[\lambda_{\Xi_1}\lambda_{\Xi_2}]
-G^{AB}v^a\left(\int d\Theta \, \lambda_{\Xi_1}\hat u_a \lambda_{\Xi_2}\right)\right)\\
&&+ \I Z_{I}^a Z_{J}^b{f_{ab}}^c u_c
- \left[X_{IA\Xi} Z_{J}^a - (I \leftrightarrow J )\right]\cT^{A}[\hat u_a \lambda_{\Xi}]
+ \I {f_{bc}}^a \left[Y_{Ia} Z_{J}^b v^c-(I\leftrightarrow J)\right]\,.
\ea
Here and in the following,
we use the Einstein summation convention, i.e.,
all repeated indices are summed over.
The action (\ref{IKKT}) can then be expanded as
\ba
S_0&=&
-Z_{I}^aZ_{J}^b{f_{ab}}^c\left[{f_{dc}}^eY_{Ie} Z_{J}^d
+\frac\I2 G^{AB}\left(\int d\Theta \, \lambda_{\Xi_1}\hat u_c \lambda_{\Xi_2}\right)
X_{IA\Xi_1} X_{JB\Xi_2}\right] \, \nn
\\
&&-\frac{1}{4}{F^{AB}}_C {F^{A'B'}}_{C'}
X_{IA\Xi_1} X_{JB\Xi_2}X_{IA'\Xi_3}X_{JB'\Xi_4}
G^{CC'}\int d\Theta \, \lambda_{\Xi_1}\lambda_{\Xi_2}\lambda_{\Xi_3}\lambda_{\Xi_4} 
\nn\\
&&-\I{F^{AB}}_C
X_{IA\Xi_1} X_{JB\Xi_2} G^{CD}X_{ID\Xi_3} Z_{J}^a
\int d\Theta \, \lambda_{\Xi_1}\lambda_{\Xi_2}\hat u_a \lambda_{\Xi_3}
\nn\\
&& +\frac12 X_{IA\Xi_1} Z_{J}^a G^{AB}\left(
X_{IB\Xi_2} Z_{J}^b
\int d\Theta \, \hat u_a\lambda_{\Xi_1}\hat u_b \lambda_{\Xi_2} -(I\leftrightarrow J)
\right) \, .
\label{S0m-explicit}
\ea

\subsection{$\hat{C}$ components as $\mathfrak{k}$-valued fields on $G/H$}

We notice that the expression above 
can be reinterpreted as an action for $\mathfrak{k}$-valued fields
living on the coset space.
Let $\mathbf{T}^A$ denotes the Lie algebra generators of $\mathfrak{k}$,
and $G^{AB} =\mathrm{Tr}  (\mathbf{T}^A \mathbf{T}^B)$
the invariant metric defined from the trace on $\mathfrak{k}$.
Let $\pj$ denotes a linear map from $\hat{C}(K; G, H)$
to $\mathfrak{k}$-valued functions on $G/H$ defined by
\begin{align}
\label{pj-def}
\pj\left(C_{A\Xi}\cT^A[\lambda_\Xi] + y_a v^a + z^{a} u_a\right)
&\equiv
C_{A\Xi} \lambda_{\Xi}(\theta) \mathbf{T}^A \, .
\end{align}
In particular,
\begin{equation}
\pj(X_I)=\sum_{\Sigma}X_{IA\Sigma} \lambda_{\Sigma}(\theta) \mathbf{T}^A =\bar{X}_I(\theta)
\end{equation}
is a $\mathfrak{k}$-valued field on $G/H$.

Furthermore, we define a map $\lb\bullet\rb$ from $\mathfrak{k}$-valued functions on $G/H$ to $\mathbb{C}$ as
\begin{align}
\lbl C^{(1)} C^{(2)} \cdots \rbl
\equiv \int d\Theta \; \mbox{Tr}\left({C}^{(1)}(\theta) \, {C}^{(2)}(\theta)\cdots\right).
\end{align}

Using the map $\pj$ and $\lb\bullet\rb$,
the action can then be more concisely expressed as
\ba
S_0&=& \left< [Z_I, Z_J],[Y^I, Z^J]\right>-\frac12 
\left<[Z_I, Z_J],v^a\right>\lb \bar X^{I} \left(\hat u_a \bar X^{J}\right) \rb \, \nn
\\
&&+\frac{1}{4}\lb
\left[\bar X_{I}, \bar X_{J}\right]\left[\bar X^{I}, \bar X^{J}\right]\rb
-Z_{J}^a\lb[\bar X_I, \bar X^J] \left(\hat u_a \bar X^I\right) \rb
\nn\\
&&+\frac12 Z_{J}^a\left(
Z^{Jb} \lb\left(\hat u_a \bar X_I\right)\left(\hat u_b \bar X^I\right)\rb
-Z^{Ib}\lb\left(\hat u_a \bar X_I\right)\left(\hat u_b \bar X^J\right)\rb
\right) \, .
\ea
We note that the terms written in the $\lb \bullet \rb$ resemble the field-theoretical action. 
On the other hand,  the ghost fields $Y_I$ and $Z_I$, being independent of the base space,
appear as non-dynamical variables.

The equations of motion are derived from the action above by variation
with respect to $Y_I$, $Z_I$ and $X_I$, respectively, as
\ba
0&=&\sum_J [[Z_I, Z_J], Z^J]\label{ZZZ} \, ,
\\
0&=&-\sum_J \left(
[[Z_I, Z_J],Y^J]-[Z_J,[Z_I,Y^J]]+[Z_J,[Z^J,Y_I]]
\right)-
\sum_{J,a} [Z_J,v^a]\lb \bar X_I, \hat u_a\bar X^J\rb
\nn\\
&&+ \sum_{J,a} v^a \lb[\bar X_I, \bar X_J],\hat u_a \bar X^J\rb
+\sum_{J,a,b} v^a \left(
\lb \hat u_a \bar X_J,\hat Z_I \bar X^J\rb -\lb \hat u_a \bar X_I, \hat Z_J \bar X^J\rb
\right)
\label{ZZY} \, ,
\\
0&=&
-\sum_J (\hat Z_J \hat Z^J \bar X_I -\hat Z_J \hat Z_I \bar X^J)\\
&&-\left([\bar X_J,\hat Z^J \bar X_I]-[\bar X_J,\hat Z_I \bar X^J]
-\hat Z_J[\bar X_I,\bar X^J] 
\right)-[[\bar X_I, \bar X_J],\bar X^J] \, ,
\label{ZZX}
\ea
where we have used the notation $\hat Z_I \equiv \sum_aZ_{I}^a \hat u_a$,
with $\hat{u}_a$ defined by eq.(\ref{uhat-def}).
A choice of the non-dynamical parameters $Y$ and $Z$
is thus constrained by eqs.(\ref{ZZZ}) and (\ref{ZZY}).
The equation for $Z$ (\ref{ZZZ}) is closed by itself,
and a different choice of the solution of $Z$ changes
the coefficients of the equation of motion for $\bar X_I(\theta)$ (\ref{ZZX}).
On the other hand, $Y$ appears only in eq.(\ref{ZZY}),
with the rest of the equations independent of $Y$.
It is a Lagrange multiplier with the only purpose of
imposing the equation of motion for $Z$ (\ref{ZZZ}).

In the following sections, 
we will omit the bar in $\bar X_I(\theta)$ as $X_I(\theta)$
for the simplicity of the notation.

\subsection{Solution for $Z$}

\paragraph{Abelian case}

When $G=U(1)^{\otimes \ell}$ and $H$ is trvial,
eq.(\ref{ZZZ}) gives no constraint on $Z$.
There is a global symmetry $O(d, \mathbb{R})$ which rotates $\cX_I$ as
$\cX_I \rightarrow \cX'_{I}=\sum_J L_{I}{}^{J} \cX_{Ja}$,
inducing a rotation on $Z_I$.
There is another global symmetry $O(\ell, \mathbb{Z})$ that rotates
$(u_a, v^a) \rightarrow 
(u'_a = \sum_b M_a{}^b u_b, v'{}^a = \sum_b v^b M^{-1}{}_b{}^a)$.
Assuming $\ell\leq d$,
the rotation symmetries allow us to set,
without loss of generality,
\ba
\label{eq:Zsolution}
Z_{Ia} = \left\{\begin{array}{ll}
L_{Ia}
\quad & (I=1,\cdots, \ell; \;\; a \leq I) \, ,
\\
0 & (I=\ell+1,\cdots, d)
\end{array}\right.
\ea
where $L_{Ia}$ encodes the modular parameters
of the $\ell$-dimensional torus $T^{\ell}$.
The action $S_0^m$ (\ref{S0m-explicit}) becomes simply
\ba\label{e:IKKT}
S_0^m = \frac{1}{4}\lb [D_I, D_J]^2\rb,
\ea
where
\begin{align}
D_I \equiv
\I Z_{I}^a\partial_{a} + X_I(\theta)
=:\left\{
\begin{array}{cl}
\I L_{I}^a\partial_{a} + A_I(\theta) \quad & (I = 1, \cdots, \ell) \, ,
\\
\Phi_I(\theta) 
&  (I = \ell+1, \cdots, d) \,.
\end{array}
\right.
\end{align}
This is the action for the Yang-Mills theory (with the gauge group $K$)
dimensionally reduced from $d$-dimensions to $T^{\ell}$.
The components of the gauge potential in the reduced dimensions
are turned into $(d-\ell)$ scalar fields $(\Phi_{\ell+1}, \cdots, \Phi_d)$
in the adjoint representation.

Our approach uses the equation of motion (\ref{ZZZ}) to derive the torus modulus $L_{Ia}$.
This equation comes from the variation of $Y_I$, the coefficient of the ghost $v^a$, which has not been considered in IKKT approaches in the past.
This is an analogue of the Higgs-like mechanism \cite{Ho:2008ei} used in the BLG-type description of M2-branes.

\paragraph{A problem for non-Abelian $G$}

Suppose $\mathfrak{g}$ is a simple Lie algebra such as $su(2)$. 
An obvious solution to eq.(\ref{ZZZ}) is
\ba
Z_{I}=Z_I u,  \qquad u \equiv \sum_a \phi^a u_a, \qquad\phi^a\in \mathbb{R} \, .
\ea
With such solutions, 
however, 
we would have a single derivative
$\hat u=\sum_a \phi^a \hat u_a$ appearing in the action,
instead of $|G|-|H|$ independent derivatives for the coset space $G/H$.
It means that we have an infinite number of states at each KK level, and
it does not produce a field theory living on $G/H$. 
Therefore, we would like to consider modifications of the action
to admit solutions of $Z$ that would lead to
a field theory on the coset space $G/H$
with its isometry $G$ as a global symmetry.

\section{Gauge theory with quadratic and cubic terms}

Motivated by the problem mentioned above for a non-Abelian group $G$,
we consider adding an extra quadratic (mass) term
and a cubic (Chern-Simons-like) term
to the
action (\ref{IKKT}).
It turns out that the modified equation of motion for $Z_I$ 
has non-trivial solutions leading to 
a field theory on $G/H$ with the isometry as a global symmetry \cite{Ishii:2008tm, Kawai:2009vb}.
Since there should be $|G|$ isometry transformation generators
realized as differential operators through the $Z_I$'s,
we must consider
$d \geq |G|$.
Using the $O(d; \mathbb{R})$ symmetry acting on the index $I$ of $\cX_I$,
we can choose these $|G|$ differential operators $Z_I$ to belong to
the first $|G|$ components $\cX_a$ ($a = 1, 2, \cdots, |G|$).
The remaining components $\cX_I$ ($I = |G| + 1, \cdots, d$)
would merely contribute more scalar fields to the model.

We shall first consider the case $d = |G|$
to focus on the components $\cX_a$ ($a = 1, 2, \cdots, |G|$),
and assume that the metric $\eta_{IJ}$
agrees with the invariant inner product $\eta_{ab}$ of $\mathfrak{g}$.
It will be straightforward to extend the result to $d > |G|$
by adding more scalar fields $\pj(\cX_{|G|+1}), \cdots, \pj(\cX_d)$ in the end.

\subsection{Modified action}
For the choice of Lie algebra $\mathfrak{g}$, we restrict ourselves to a tensor product of semi-simple Lie algebras and an abelian algebra, namely
$\mathfrak{g}=(\oplus_{\iota=1}^\mathfrak{s} \mathfrak{g}_\iota)\oplus \mathfrak{g}'$ where $\mathfrak{g_\iota}$ ($\iota=1,\cdots, \mathfrak{s}$) are simple Lie algebras, and $\mathfrak{g}'$ is abelian. It is clear that the parameters of the action, which will be considered in the following, can be separately chosen for each factor. Since the analysis for the abelian part will be the same as in the previous section, we will focus on one of the simple Lie algebras $\mathfrak{g}_\iota$ and omit the index $\iota$.

The action $S_0$ (\ref{IKKT}) is modified by quadratic and cubic terms as
\begin{equation}
S=S_0+S_1+S_2 \, ,
\label{action-modified}
\end{equation}
with
\begin{align}
S_1 &= R^2 h^{\mathfrak{g}}_{a b} \left< \cX^{a}, \cX^{b} \right> \, ,
\label{eq:massterm}
\\
S_2 &= \I \lambda f_{a b c} \left< \cX^{a}, [\cX^{b}, \cX^{c}] \right> \, .
\label{eq:cubicterm}
\end{align}
where 
\begin{align}
- f_{ac}{}^{d} f_{bd}{}^{c}=h^{\mathfrak{g}}_{ab}
\label{ff-h}
\end{align}
is the Killing form of $\mathfrak{g}$.
For the abelian factor, $S_1$ should vanish for the consistency of equation of motion and $S_2$ vanishes since the structure constants are absent.
For the new action,
the equation of motion for $Z_I$ (\ref{ZZZ}) is
\ba
[[Z_{a},Z_{b}],Z^{b} ]=2R_\iota^2 h^{\mathfrak{g}}_{ab} Z^b+3{\rm i}\lambda f_{a b c}[Z^{b},Z^{c}] \, .
\label{Z-new-eq}
\ea
Using eq.(\ref{ff-h}),
we write a solution in the form
\ba
\label{eq:Zsolution2}
Z_{a}=L u_{a} \, .
\ea
With this ansatz, the equation (\ref{Z-new-eq}) becomes
\ba
\label{eq:Lequation}
L^2+3\lambda L-2R^2=0 \, .
\ea
%
For an Abelian algebra, the corresponding $L$ parameters remain arbitrary, as in the previous section. For the semi-simple part, we shall assume that $L \in \mathbb{R}$,
but it is not yet clear whether we need $R^2 > 0$
for a real mass
as $S_0$ and $S_2$ may also contribute to
the mass term of the scalar fields.

 
Substituting eq.(\ref{eq:Zsolution2}) into the action (\ref{action-modified}),
we obtain 
\begin{equation}
\label{eq:u1action}
\begin{split}
S=&
\frac{1}{4}\lbg\left(\hat{Z}_a X_b-\hat{Z}_b X_a+[X_a,X_b]\right)^2\rbg
-\left(\frac{1}{2}+3\frac{\lambda}{L}\right)
\lbl X_a[\hat{Z}^a,\hat{Z}^b]X_b\rbl + R^2 h^{\mathfrak{g}}_{ab}\lb X^a X^b\rb 
+{\rm i}\lambda f_{abc}\lb [X^a,X^b]\, X^c\rb
\\
=&
\frac{1}{4}\lbl\left(\hat{Z}_a X_b-\hat{Z}_bX_a+[X_a,X_b]-{\rm i}Lf_{abc}X^c\right)^2\rbl
\\&
+\frac{3}{2}(L+2\lambda)\left(-{\rm i}f_{abc}\lb X^a \, \hat{Z}^cX^b\rb
+\frac{Lh^G_{ab}}{2}\lb X^a X^b\rb+\frac{2{\rm i}}{3}f_{abc}\lb X^a X^b X^c\rb\right),
\end{split}
\end{equation}
where we have used eq.(\ref{eq:Lequation}) to replace $R$ by $L$ and $\lambda$.
Eq.(\ref{eq:Zsolution2}) implies that $\hat{Z}_a = L\hat{u}_a$,
and the action becomes
\ba
S&=&S_{\rm kinetic}+S_{CS},
\label{S=Sk+SCS}
\\
\label{eq:kineticaction}
S_{\rm kinetic}&\equiv&
\frac{1}{4}
\lb\left(\hat{Z}_a X_b-\hat{Z}_bX_a+[X_a,X_b]-{\rm i}Lf_{abc}X^c\right)^2\rb
\nonumber\\
&=&\frac{L^4}{4}
\int d\Theta \; \mbox{Tr}
\left[\left(\hat{u}_a\hat{X}_b-\hat{u}_b\hat{X}_a
+[\hat{X}_a,\hat{X}_b]-{\rm i}f_{abc}\hat{X}^c\right)^2\right] \, ,
\\
\label{eq:CSlikeaction}
S_{CS}&\equiv&
\frac{3}{2}(L+2\lambda)\left(-{\rm i}f_{abc}
\lb X^a \, \hat{Z}^c X^b\rb
+\frac{Lh^{\mathfrak{g}}_{ab}}{2}\lb X^a X^b\rb
+\frac{2{\rm i}}{3}f_{abc}\lb X^a X^b X^c\rb\right)
\nonumber\\
&=&\frac{3}{2}(L+2\lambda)L^3
\int d\Theta \; \mbox{Tr}
\left[
-{\rm i}f^{abc}
\hat{X}_a \, \hat{u}_c\hat{X}_b
+\frac{h^\mathfrak{g}_{ab}}{2}
\hat{X}^a \hat{X}^b
+\frac{2{\rm i}}{3}f^{abc}
\hat{X}_a \tilde{X}_b \hat{X}_c
\right]
\nonumber\\
&=&\frac{3\rm i}{4}f^{abc}(L+2\lambda)L^3 \int d\Theta \;
\mbox{Tr}\left[\hat{X}_a\left(\hat{u}_b \hat{X}_c - \hat{u}_c \hat{X}_b
+ [\hat{X}_b,\hat{X}_c] - {\rm i} f_{dbc}\hat{X}_d\right)
-\frac{1}{3}[\hat{X}_a,\hat{X}_b] \hat{X}_c\right] \, ,
\label{eq:CSlikeaction3}
\ea
where $\hat{X}_a$ is a $\mathfrak{k}$-valued field
on $G/H$ defined by (see eq.(\ref{pj-def}))
\begin{align}
\hat{X}_a \equiv \frac{1}{L} \, X_{a}(\theta) \, .
\end{align}

\subsection{Gauge invariance}
Before $Z_I$ takes a specific solution,
the action (\ref{action-modified}) is manifestly invariant under
the transformation
\begin{align}
\delta\cX_a = [\cX_a, {\cal E}]
\label{dcXa}
\end{align}
for any ${\cal E} \in \hat{C}(K; G, H)$.
For
\begin{align}
{\cal E} = \epsilon + \kappa + \xi
= \epsilon_{A\Xi}\cT^A[\lambda_{\Xi}]
+ \kappa_b v^b + \xi^b u_b \, ,
\end{align}
the components of $\cX_a$ transform
for the background configuration (\ref{eq:Zsolution2}) as
\begin{align}
\delta X_a &=
\I F^{AB}{}_CX_{aA\Xi_1}\epsilon_{B\Xi_2}\cT^C[\lambda_{\Xi_1}\lambda_{\Xi_2}]
+ L \epsilon_{A\Xi} \cT^A[\hat{u}_a\lambda_{\Xi}]
- \xi^b X_{aA\Xi}\cT^A[\hat{u}_b\lambda_{\Xi}] \, ,
\label{dXa}
\\
\delta Y_a &= \left(-G^{AB} X_{aA\Xi_1} \epsilon_{B\Xi_2}
\int d\Theta \, \lambda_{\Xi_1} \hat{u}_b \lambda_{\Xi_2}
- \I L f_{ab}{}^c \kappa_c 
+ \I f_{cb}{}^d Y_{ad}\xi^c\right) v^b \, ,
\\
\delta Z_a &= \I L f_{ab}{}^c \xi^b u_c \, .
\end{align}
The background configuration of $Z_a$ breaks the $\hat{C}$-symmetry
to the partial symmetry constrained by $\xi^a = 0$.
(Recall that $Y_a$ is decoupled from other fields
as a Lagrange multipler,
so we do not need to demand that $\delta Y_a = 0$.)
The residual symmetry transformation is thus equivalent to
\ba
\label{eq:gaugesym2}
\delta \hat{X}_a=\hat{u}_a\hat\epsilon+[\hat{X}_a,\hat\epsilon] \, ,
\ea
where $\hat{\epsilon} \equiv \pj({\cal E})$.
Hence, this action defines a non-Abelian gauge theory
with the gauge group $K$ on the coset space $G/H$.
Under this gauge transformation,
the actions $S_{\rm kinetic}$ and $S_{CS}$ are individually invariant.

In order to derive a more standard action for the gauge theory, we need to decompose $X_a(\theta)$ into the gauge field $A_\mu(\theta)$ and the scalar field $\Phi_i$, on which we will focus in the following sections.

We note that the action we obtained are the same
as those given in Ref.\cite{Kawai:2010sf} when $H$ is trivial, namely for the group manifold. 
While the authors of Ref.\cite{Kawai:2010sf} used the dimensional reduction to obtain an action on $G/H$, we applied a purely algebraic method. Ref.\cite{Kawai:2010sf} introduced a ``minimal action," which contains only the gauge potential $A_\mu$. The derivation of such an action is not obvious in our purely algebraic framework.

\hide{
As the gauge invariance of $S_{\rm kinetic}$ survives
as the residual gauge symmetry,
we check it explicitly as follows.
First,
\begin{equation}
\label{eq:k-invariance}
\begin{split}
\delta S_{\rm kinetic}&=\frac{1}{2}\lb\left[(\hat{Z}_a X_b-\hat{Z}_bX_a+[X_a,X_b]-{\rm i}f_{abc}X_c),\epsilon\right]
\left(\hat{Z}_aX_b-\hat{Z}_bX_a+[X_a,X_b]-{\rm i}f_{abc}X_c\right)\rb\\
&=0
\end{split}
\end{equation}
To see the invariance of $S_{CS}$, it is convenient to rewrite it into
\ba
\label{eq:CSlikeaction2}
S_{CS}=\frac{3\rm i}{4}f_{abc}(L+2\lambda)
\left(\lb X_a\left(\hat{Z}_bX_c-\hat{Z}_cX_b
+[X_b,X_c]-{\rm i}Lf_{dbc}X_d\right)\rb
-\frac{1}{3}\lb [X_a,X_b] X_c\rb\right) \, .
\ea
Then the variation is given by
\ba
\delta S_{CS}&=&
\frac{3\rm i}{4}f_{abc}(L+2\lambda)
\Bigl(\lb\hat{Z}_a\epsilon,\ \hat{Z}_bX_c-\hat{Z}_cX_b+[X_b,X_c]
-{\rm i}Lf_{dbc}X_d\rb 
-\lb [X_b,X_c]\hat{Z}_a\epsilon+[X_a,\epsilon]\rb\Bigr)
\nonumber \\
&=&\frac{3\rm i}{4}f_{abc}(L+2\lambda)\lb\hat{Z}_a\epsilon
\left(\hat{Z}_bX_c-\hat{Z}_cX_b-{\rm i}Lf_{dcb}X_Md\right)\rb
\nonumber \\
&=&-\frac{3\rm i}{4}f_{abc}(L+2\lambda)
\lb\epsilon \left([\hat{Z}_a,\hat{Z}_b]X_c - {\rm i}Lf_{dbc}\hat{Z}_aX_d\right)\rb
\nonumber \\
&=&0\ ,
\ea
where we use the equation
\ba
f_{abc}\lb [X_b,X_c][X_a,\epsilon]\rb=f_{abc}\lb [[X_b,X_c],X_a]\epsilon\rb=0
\ea
and integration by part.
}

The Chern-Simons-like action (\ref{eq:CSlikeaction3}) is defined on the coset space whose dimension
can be equal to or larger than
three. It will take the form $\int d\Theta \, C_{\mu\nu\rho}(\theta) \chi^{\mu\nu\rho}$ where $C_{\mu\nu\rho}$ is the three-form induced from the structure constant $f_{abc}$ of $G$ via the vielbein, and $\chi^{\mu\nu\rho}$ is the Chern-Simons term with the gauge group $K$.



\section{Reduction to the conventional gauge theory on the coset: $G/H=SU(2)/U(1)=S^2$}
\label{sec-S2}

To be explicit, we first consider the $S^2$ case where we have a standard description in terms of the polar coordinates.
It corresponds to a fixed coset representative in appendix \ref{cosetS2}.

We introduce the orthogonal basis as
\ba
e_a{}^r=e_r{}^a=\vec{e}_r&=& (\sin\theta\cos\varphi,\sin\theta\sin\varphi,\cos\theta)\\
e_a{}^\theta=e_\theta{}^a =\vec e_\theta&=& (\sin\varphi,-\cos\varphi,0)\\
e_a{}^\varphi \sin\theta=e_\varphi{}^a(\sin\theta)^{-1}=\vec e_\varphi&=& (\cos\theta\cos\varphi, \cos\theta\sin\varphi,-\sin\theta)
\ea
These are the orthogonal basis satisfying $\delta^{ab}e_a{}^\mu e_b{}^\nu=g^{\mu\nu}$, 
$\delta_{ab} \, e_\mu{}^a e_\nu{}^b=g_{\mu\nu}$ with a diagonal metric $g_{rr}=g_{\theta\theta}=1$, $g_{\varphi\varphi}=\sin\theta$.
$\vec e_\mu$ is set to be orthonormal.
We can rewrite eqs.(\ref{u1}, \ref{u2}, \ref{u3}) as
\ba
\hat u_a= \I\left(e_a{}^\theta\partial_\theta+e_a{}^\varphi\partial_\varphi\right).
\label{eq:uacomponent}
\ea
This expression and the gauge symmetry (\ref{eq:gaugesym2}) imply that  the field $\hat{X}$ should be decomposed in the following way:
\begin{equation}
\label{eq:S2component}
\begin{split}
&\hat{X}_a=e_a{}^\theta A_\theta+e_a{}^\varphi A_\varphi+e_a{}^r\Phi \, .\\
\end{split}
\end{equation}
For the computation below, it is convenient to rewrite (\ref{eq:uacomponent}) and (\ref{eq:S2component}) in the vector notation as follows:
\ba
&&\vec{u}=\I \left(\vec{e}_\theta\partial_\theta+\frac{\vec{e}_\varphi}{\sin\theta}\partial_\varphi\right),\\
&&\vec{X}=\vec{e}_r\Phi+\vec{e}_\varphi \frac{A_\varphi}{\sin\theta}+\vec{e}_\theta A_\theta.\label{eq:S2component2}
\ea
We also introduce the field strength and the covariant derivative;
\ba
&&F_{\theta\varphi}=\partial_\theta A_\varphi-\partial_\varphi A_\theta-\I[A_\theta,A_\varphi],\\
&&D_{\theta}\Phi=\partial_\theta\Phi-\I[A_\theta,\Phi],\qquad D_{\varphi}\Phi=\partial_\varphi\Phi-\I[A_\varphi,\Phi].
\ea

\paragraph{The action and the separation of variables}
For the $S^2$ case, the actions (\ref{eq:kineticaction}) and (\ref{eq:CSlikeaction}) are simplified to
\begin{equation}
\begin{split}
\label{eq:S2action}
S_{\rm kinetic}&=  \frac{L^4}{4}\sum_{a,b} \int d\Theta \ {\rm Tr}\left(\hat u_a \hat{X}_b-\hat u_b \hat{X}_a+[\hat{X}_a,\hat{X}_b]-\I\epsilon_{abc}\hat{X}_c\right)^2
\\
&=\frac{L^4}{2}\int d\Theta \ {\rm Tr} \left(\vec u\times \vec X+\vec X\times\vec X-\I \vec{X}\right)^2,
\end{split}
\end{equation}
\begin{equation}
\begin{split}
\label{eq:S2CSaction}
S_{CS}&=\frac{3\I(L+2\lambda)L^3}{2}\int d\Theta \ {\rm Tr}\left(\vec{X}\cdot(\vec u\times \vec X+\vec X\times\vec X-\I \vec{X})-\frac13\vec{X}\cdot(\vec{X}\times\vec{X})\right)
\end{split}
\end{equation}
in the vector notation.
To express them by the components (\ref{eq:S2component2}), we need the following formulae,
\begin{equation}
\begin{split}
&\partial_{\theta}\vec{e}_r=\vec{e}_{\varphi},\qquad \partial_{\varphi}\vec{e}_r=-\vec{e}_{\theta}\sin\theta,\qquad \partial_{\theta}\vec{e}_\theta=0 \, ,\\
& \partial_{\varphi}\vec{e}_\theta=\vec{e}_{\varphi}\cos\theta+\vec{e}_r\sin\theta \, ,\qquad\partial_{\theta}\vec{e}_\varphi=-\vec{e}_r,\qquad\partial_{\varphi}\vec{e}_\varphi=-\vec{e}_\theta\cos\theta \, ,
\end{split}
\end{equation}
\begin{equation}
\begin{split}
\vec{u}\times\vec{X}+\vec X\times\vec X-\I\vec{X}=&{\rm i}\left(\frac{1}{\sin\theta}F_{\theta\varphi}+\Phi\right)\vec{e}_r+\frac{\I}{\sin\theta}(D_\varphi\Phi)\vec{e}_\theta-{\rm i}(D_\theta\Phi)\vec{e}_\varphi \, .
\end{split}
\end{equation}
Using these formulae, we can  transform the actions into the following form,
\ba
&&S_{\rm kinetic}=-\frac{L^4}{2}\int d\Theta \ {\rm Tr}\left(\frac{1}{\sin^2\theta}F_{\theta\varphi}^2+(D_\theta\Phi)^2+\frac{1}{\sin^2\theta}(D_\varphi\Phi)^2+\Phi^2+\frac{2}{\sin\theta}F_{\theta\varphi}\Phi \right) \, ,\hspace{90pt}
\ea
\begin{equation}
\begin{split}
S_{CS}&=-\frac{3(L+2\lambda)L^3}{2}\int d\Theta \ {\rm Tr}\Biggl(\Phi\left(\frac{F_{\theta\varphi}}{\sin\theta}+\Phi\right)+\frac{1}{\sin\theta}\bigl(A_\theta D_\varphi\Phi-A_\varphi D_\theta\Phi\bigr)+\frac{\I}{\sin\theta}\Phi[A_\theta,A_\varphi]\Biggr)\\
&=-\frac{3(L+2\lambda)L^3}{2}\int d\Theta \ {\rm Tr}\Biggl(\frac{2}{\sin\theta}F_{\theta\varphi}\Phi +\Phi^2\Biggr) \, ,
\end{split}
\end{equation}
where we use integration by part in the last line. Combining them, we have 
\ba
S=-\frac{L^4}{2}\int d\Theta \ {\rm Tr}\Biggl(\frac{1}{\sin^2\theta}F_{\theta\varphi}^2+(D_\theta\Phi)^2+\frac{1}{\sin^2\theta}(D_\varphi\Phi)^2+2\left(2+\frac{3\lambda}{L}\right)\Bigl(\Phi^2+\frac{2}{\sin\theta} F_{\theta\varphi}\Phi\Bigr)\Biggr) \, .
\ea
This action has the standard Yang-Mills term
with the gauge field and a massive scalar field living on the $2$-sphere.
The scalar field is in the adjoint representation
with a non-minimal coupling to the gauge field.
The Chern-Simons term written in terms of gauge fields is absent because
it can only exist for a base space of 3 or higher dimensions.

\section{Decomposition of $\hat{X}_a$ into gauge potential and scalar field}


As a generalization of the example studied in the previous section,
we expect that $\hat{X}_a$ is in general a linear combination of 
the gauge potential $A_{\mu}$ ($\mu = 1, 2, \cdots, |G| - |H|$)
and scalar fields $\Phi_i$ ($i = 1, 2, \cdots, |H|$).
For generic $G$ and $H \subset G$,
let
\begin{align}
\hat{X}_a = e_a{}^{\mu} A_{\mu} + e_a{}^i \Phi_i \, ,
\label{X-comp}
\end{align}
where both $A_{\mu}$ and $\Phi_i$ are $\mathfrak{k}$-valued 
fields on the coset space $G/H$.
For a dimensional reduction of $G \rightarrow G/H$,
the bases $e_a{}^{\mu}(\theta)$ and $e_a{}^i(\theta)$
provide a local decomposition of the tangent space of the group $G$
into the tangent space of the base space $G/H$
and directions along the $H$-fibers.

\subsection{Algebraic properties of basis}

The basis $e_a{}^{\mu}$ is naturally defined by
\begin{align}\label{Cbasis-1}
{e_a}^\mu(\theta)& \equiv -\I\, {u_a}^\mu(\theta) 
= {D_a}^\alpha(\sigma(\theta)) {V_\alpha}^\mu(\theta) \, ,
\end{align}
so that $A_{\mu}$ appears in the combination $(\partial_{\mu} - \I A_{\mu})$
of a covariant derivative in $\cX_I$.
The basis $e_a{}^i$ is chosen to be
\begin{align}\label{Cbasis-2}
e_a{}^i(\theta) \equiv {D_a}^i(\sigma(\theta)) \, ,
\end{align}
so that it is orthogonal to $e_a{}^{\mu}$.
The inverse of the bases $(e_a{}^{\mu}, e_a{}^i)$ is given by
\begin{align}
{e_\mu}^a(\theta) = {D^a}_\beta(\sigma(\theta)) V^\beta{}_\mu(\theta)
\, ,
\qquad
{e_i}^a(\theta) = {D^a}_i(\sigma(\theta)) 
\, . 
\end{align}
They satisfy
\begin{align}
& \eta^{ab} \ {e_a}^\mu  {e_b}^i=0 \, , 
\quad
\eta_{ab} \ {e_\mu}^a{e_i}^b=0 \, ,
\\
& \eta^{ab} \ {e_a}^i  {e_b}^j=\eta^{ij} \, ,
\quad
\eta_{ab} \ {e_i}^a{e_j}^b=\eta_{ij} \, ,
\label{h-metric}
\\
& {e_i}^a {e_a}^j =\delta_i^j \, ,
\quad 
{e_i}^a {e_a}^\mu ={e_\mu}^a {e_a}^i=0 \, ,
\quad
{e_\mu}^a {e_a}^\nu=\delta_\mu^\nu \, ,
\\
& {e_a}^i {e_i}^b +{e_a}^\mu {e_\mu}^b=\delta_a^b \, .
\end{align}
Eq.(\ref{h-metric}) indicates that $e_a{}^i$
is an orthonormal basis for the field space of $\Phi_i$.

The metric $g_{\mu\nu}$ and its inverse (\ref{metric-g}) on the coset space $G/H$ 
for the coordinates $\theta$ are related to the vielbein $e_a{}^{\mu}$
via the following relations as usual:
\begin{align}
g_{\mu\nu}(\theta) &= \eta_{ab} \, e_{\mu}{}^a(\theta) e_{\nu}{}^b(\theta) \, ,
\\
g^{\mu\nu}(\theta) &= \eta^{ab} \, e_a{}^{\mu}(\theta) e_b{}^{\nu}(\theta) \, .
\end{align}
The indices $\mu$, $\nu$ can be raised or lowered using $g^{\mu\nu}$ and $g_{\mu\nu}$.

\subsection{Transformation properties of basis}

Under the $H$-gauge transformation (\ref{h-gauge}),
the basis $({e_a}^\mu, {e_a}^i)$ transform as
\begin{align}
{e_a}^\mu
\quad &\rightarrow \quad
{\tilde e_a}{}^\mu ={D_a}^\beta(\sigma) {D_\beta}^\gamma(\tilde{h}) 
{\bar{D}}_\gamma{}^\alpha(\tilde{h}) {V_\alpha}^\mu
={D_a}^\alpha(\sigma) {V_\alpha}^\mu={e_a}^\mu \, ,
\label{h-gauge3}
\\
{e_a}^i
\quad &\rightarrow \quad 
{\tilde e}_a{}^i ={D_a}^i (\sigma \tilde{h})
={D_a}^b(\sigma){D_b}^i(\tilde{h})
={D_a}^j(\sigma){D_j}^i(\tilde{h})
={e_a}^j{D_j}^i (\tilde{h}) \, .
\label{h-gauge2}
\end{align}
The equations (\ref{h-gauge3}, \ref{h-gauge2}) imply
the transformation of the component fields:
\ba
\tilde \Phi_i = {D(h^{-1})_i}^j \Phi_j\,,\quad
\tilde A_\mu = A_\mu\,.
\ea

Under the infinitesimal isometry generated by $\hat{u}_a$,
the basis behaves as
\begin{align}
& {u_a}^\mu\partial_\mu {e_b}^\nu-{u_b}^\mu\partial_\mu {e_a}^\nu
=\I {f_{ab}}^c {e_c}^\nu \, ,
\label{umu-enu}
\\
& {u_a}^\mu\partial_\mu {e_b}^i
=\I {f_{a b}}^c {e_c}^i -\I {f_{jk}}^i \left({e_a}^j- {e_a}^\mu \Omega_\mu^j \right) {e_b}^k
=\I {f_{a b}}^c {e_c}^i -\I {f_{jk}}^i {\Lambda_a}^j {e_b}^k \, .
\label{umu-ei}
\end{align}
The first line is the immediate consequence of eq.(\ref{upu}).
The second line needs more explanation. 
We calculate (using ${D_b}^i T^b={(D^T)^i}_b T^b=\sigma T^i \sigma^{-1}$, ${D_a}^b={D_a}^b(\sigma)$)
as follows:
\begin{align*}
{e_a}^\mu\partial_\mu ({e_b}^i T^b) &= {D_a}^\alpha {V_\alpha}^\mu \partial_\mu \left(\sigma T^i
 \sigma^{-1}\right)
 \\
 &= {D_a}^\alpha {V_\alpha}^\mu \sigma\left(\left[\sigma^{-1}\partial_\mu \sigma, T^i\right]\right)\sigma^{-1}
 \\
 &=\I \,{D_a}^\alpha {V_\alpha}^\mu \sigma\left(\left[V_\mu^\beta T_\beta+\Omega_\mu^j T_j, T^i\right]\right)\sigma^{-1}
 \\
 &=\I\, {D_a}^\alpha \sigma\left([T_\alpha, T^i]+V_\alpha^\mu \Omega_\mu^j [T_j, T^i]
 \right)\sigma^{-1}
 \\
 &={f_{\alpha \beta}}^i {D_a}^\alpha {D_b}^\beta T^b+ {f_{jk}}^i {e_a}^\mu \Omega_\mu^j {D_b}^k T^b \, ,
\end{align*}
which implies
\begin{equation}
{e_a}^\mu\partial_\mu {e_b}^i=  {f_{\alpha \beta}}^i {D_a}^\alpha {D_b}^\beta + {f_{jk}}^i {e_a}^\mu \Omega_\mu^j {e_b}^k\,.
\end{equation}
To proceed further, we use
\begin{equation}
{f_{abc}}{D_d}^a{D_e}^b{D_f}^c=f_{def} \, ,
\quad \mbox{and}\quad 
{f_{ab}}^c{D_d}^a{D_e}^b={f_{de}}^f{D_f}^c \, ,
\end{equation}
which come from $\left[ \tilde T_a, \tilde T_b\right]=\I {f_{ab}}^c \tilde T_c$ with $\tilde T_a=\sigma^{-1} T_a \sigma$. 
It follows that
\begin{align*}
{f_{\alpha \beta}}^i {D_a}^\alpha {D_b}^\beta&={f_{c d}}^i {D_a}^c {D_b}^d-{f_{jk}}^i {D_a}^j {D_b}^k\\
&= {f_{a b}}^c {e_c}^i -{f_{jk}}^i {e_a}^j {e_b}^k \, ,
\end{align*}
which leads to eq.(\ref{umu-ei}).

The first term on the right-hand side of eq.(\ref{umu-ei})
is an analogue of eq.(\ref{umu-enu}),
and the second term gives a correction,
which vanishes when $H$ is abelian (${f_{ij}}^k=0$). 
The appearance of the $H$-connection $\Omega_\mu^i$ is necessary
since ${e_a}^i$ transforms in the adjoint representation of the $H$-gauge symmetry.
As we have seen in eq.\eqref{Omega-field},
whenever $f_{\alpha\beta}{}^i \neq 0$,
the $H$-field strength is non-zero,
and the $H$-connection cannot be gauged away.

The $H$-covariant derivative on $e_a{}^i$,
which transforms in the adjoint representation for the $H$-gauge transformation,
is
\begin{equation}
\label{eq:H-covariant}
{(\nabla_\mu e)_a}^i = 
\partial_\mu {e_a}^i - {f_{jk}}^i {\Omega_\mu}^j {e_a}^k \, .
\end{equation}
In terms of the covariant derivative, 
eq.(\ref{umu-ei}) is turned into
\begin{equation}
\label{eq:derivative-ei}
{e_a}^\mu{\nabla_\mu e_b}^i= {f_{a b}}^c {e_c}^i - {f_{jk}}^i {e_a}^j{e_b}^k \, .
\end{equation}
Similarly,
the covariant derivative $\nabla_\mu$ on $\Phi_i$
that respects both $K$- and $H$-gauge symmetries is defined as
\ba
\nabla_\mu\Phi_i=D_\mu \Phi_i -{f_{ij}}^{k} {\Omega_\mu}^j \Phi_k \, .
\ea

\section{Gauge theory on coset space from 0-dimension}

\hide{
{\color{red} In this section, we adopt the following notation,
\ba
&&-ie_a^{\mu}=Lu_a^{\mu},\\
&&(\ref{eq:H-covariant})\to{(\nabla_\mu e)_a}^i= \partial_\mu {e_a}^i -{f_{jk}}^i {\Omega_\mu}^j {e_a}^k\,\\
&&\ref{eq:derivative-ei}\to{e_a}^\mu{\nabla_\mu e_b}^i=L(-{f_{a b}}^c {e_c}^i + {f_{jk}}^i {e_a}^j{e_b}^k)
\ea
$\int d\Theta {\rm Tr}$ is omitted for now. 

PM: So I have made the following replacements to
put everything in this section to be in the same notation
as the sections above:
\begin{align}
e_a{}^{\mu} &\rightarrow - L e_a{}^{\mu} \\
e_a{}^i &\rightarrow e_a{}^i \\
A_{\mu} &\rightarrow - A_{\mu} \\
\Phi_i &\rightarrow L \Phi_i
\end{align}
}
}

In this section, 
we write down the action (\ref{S=Sk+SCS}) explicitly
in terms of the gauge potential $A_{\mu}$ and scalar fields $\Phi_i$
according to the decomposition of $\hat{X}_a$ (\ref{X-comp}) described
in the previous section.

We first consider the kinetic part (\ref{eq:kineticaction}).
By substituting the component expression (\ref{X-comp}) of $X_I$
and defining the field strength $F_{\mu\nu}$ by
\begin{align}
F_{\mu\nu} \equiv \partial_{\mu}A_{\nu} - \partial_{\nu}A_{\mu} - \I [A_{\mu}, A_{\nu}] \, ,
\end{align}
we have
\begin{align}
&\hat{Z}_a X_b-\hat{Z}_b X_a+[X_a, X_b] - \I Lf_{ab}{}^c X_c
\nonumber \\
&= \I L^2 \biggl(
e_a{}^{\mu} e_b{}^{\nu} F_{\mu\nu}
+ \left(e_a{}^{\mu}e_b{}^i-e_b{}^{\mu}e_a{}^i\right)D_{\mu}\Phi_i
- \I e_a{}^i e_b{}^j [\Phi_i,\Phi_j]
+ (e_a{}^{\mu}\partial_\mu e_b{}^i
-e_b{}^{\mu}\partial_\mu e_a{}^i
- f_{ab}{}^c e_c{}^i)\Phi_i\biggr)
\nonumber \\
&= \I L^2 \biggl(e_a{}^{\mu}e_b{}^{\nu}F_{\mu\nu}
+ \left(e_a{}^{\mu}e_b{}^i-e_b{}^{\mu}e_a{}^i\right)\nabla_{\mu}\Phi_i
- \I e_a{}^i e_b{}^j [\Phi_i,\Phi_j]
+(e_a{}^{\mu}\nabla_\mu e_b{}^i - e_b{}^{\mu}\nabla_\mu e_a{}^i 
- f_{ab}{}^c e_c{}^i)\Phi_i\biggr)
\nonumber \\
&= \I L^2 \biggl(e_a{}^{\mu}e_b{}^{\nu}F_{\mu\nu}
+ \left(e_a{}^{\mu}e_b{}^i-e_b{}^{\mu}e_a{}^i\right)\nabla_{\mu}\Phi_i
- \I e_a{}^i e_b{}^j [\Phi_i,\Phi_j]
+ (f_{ab}{}^c e_c{}^i - 2f_{jk}{}^{i}e_a{}^je_b{}^k)\Phi_i\biggr) \, .
\label{eq:calFPhi2}
\end{align}
We note that ordinary derivatives are replaced by
covariant derivatives on two terms
in the third line but their changes cancel.
In the fourth line,
we use eq.(\ref{eq:derivative-ei}).
Using eq.(\ref{eq:calFPhi2}), 
we derive
\begin{equation}
\label{eq:kineticcomponent}
\begin{split}
&S_{\rm kinetic}\\
&=\ -\frac{1}{4} L^4 \int d\Theta \;
\mbox{Tr}\left[F_{\mu\nu}F^{\mu\nu}
+ 2\nabla_\mu\Phi_i\ \nabla^\mu\Phi^i
- [\Phi_i,\Phi_j]^2
+ h^{\mathfrak{g}}_{ij}\Phi^i\Phi^j
+ 2f^{abc}e_a{}^\mu e_b{}^\nu e_c{}^i F_{\mu\nu}\Phi_i
+ 2\I f^{ijk}\Phi_i[\Phi_j,\Phi_k]\right].
\end{split}
\end{equation}

Next, 
we consider the CS-like term $S_{\rm CS}$. 
For later convenience,
we define a notation
\ba
B_{ab}^i \equiv -{f_{ab}}^c e_c{}^i + 2{f_{jk}}^i e_a{}^j e_b{}^k \, .
\ea
Using eq.(\ref{eq:CSlikeaction3}) for $S_{\rm CS}$,
due to the contraction of indices,
some of the terms vanish due to the identities
\ba
&&f^{abc} e_a{}^{\mu}e_b{}^ie_c{}^j=f^{abc}{D_a}^\alpha {V_\alpha}^\mu {D_b}^i {D_c}^j=f^{\alpha i j}{V_\alpha}^\mu=0 \, ,
\\
&&f^{abc}B_{ab}^i e_c{}^{\mu}=
-h^{\mathfrak{g}ab} e_a{}^i e_b{}^\mu + 2f^{abc} {f_{jk}}^i e_a{}^j e_b{}^k e_c{}^\mu=0 \, .
\ea
We can also use the relation
\ba
f^{abc}B_{ab}^i e_c{}^{l}=
-f^{abc}f_{ab}{}^d e_c{}^l e_d{}^i + 2f^{abc} e_a{}^j e_b{}^k e_c{}^l f_{jk}{}^i
=-h^{\mathfrak{g}}_{il}+2h^{\mathfrak{h}}_{il} \, ,
\ea
where $h^{\mathfrak{h}}$ is defined analogous to $h^{\mathfrak{g}}_{ab}$ (\ref{ff-h}) by
\begin{align}
f_{ij}{}^{k}f_{lk}{}^{j}=- h^{\mathfrak{h}}_{il} \, .
\label{ff-h2}
\end{align}
$S_{CS}$ (\ref{eq:CSlikeaction3}) can then be evaluated as
\begin{align}
S_{CS}=&\frac{3(L+2\lambda)L^3}{4} \int d\Theta \;
\mbox{Tr}\Biggl[
f^{abc}\biggl(
- e_a{}^\mu e_b{}^\nu e_c{}^\rho
(F_{\mu\nu}A_\rho + \frac{\I}{3}[A_\mu,A_\nu]A_\rho)
- 2 e_a{}^\mu e_b{}^i e_c{}^\nu A_\nu \nabla_\mu\Phi_i
\biggr.\Biggr.
\nonumber \\
&\hspace{110pt}\Biggl.\biggl.
- e_a{}^\mu e_b{}^\nu e_c{}^i F_{\mu\nu} \Phi_i
- \I e_a^\mu e_b^\nu e_c^i[A_\mu,A_\nu]\Phi_i\biggr)
+ \frac{2}{3}\I f^{ijk}[\Phi_i,\Phi_j]\Phi_k
+ (2h^{\mathfrak{h}}_{ij}-h^{\mathfrak{g}}_{ij}) \Phi^i \Phi^j
\Biggr] \, .
\label{eq:CScomponent}
\end{align}
The first term is the CS-like term discussed in Ref.\cite{Kawai:2010sf}. 
To express the remaining terms in a gauge-invariant way,
we implement integration by part on the following term:
\begin{align}
&2f^{abc}e_a{}^{\mu} e_b{}^i e_c{}^\nu  A_\nu\nabla_\mu\Phi_i
\nonumber \\
\simeq\ &2f^{abc}\left(-e_a{}^\mu\left[\partial_\mu(e_b{}^ie_c{}^\nu A_\nu)\Phi_i\right]
- \I e_a{}^\mu e_b{}^i e_c{}^\nu A_\nu[A_\mu,\Phi_i]
- e_a{}^\mu e_b{}^i e_c{}^\nu A_\nu f_{jki}\Omega_\mu^j\Phi^k\right)
\nonumber \\
=\ &f^{abc}\left(-e_a{}^\mu e_b{}^i e_c{}^\nu(\partial_\mu A_\nu-\partial_\nu A_\mu)\Phi_i
- e_b{}^i(e_a{}^\mu\partial_\mu e_c{}^\nu - e_c{}^\mu\partial_\mu e_a{}^\nu)A_\nu\Phi_i
- (e_a{}^\mu\partial_\mu e_b{}^i-e_b{}^\mu\partial_\mu e_a{}^i)e_c{}^\nu A_\nu\Phi_i\right.
\nonumber \\
&\hspace{2.1em}\left.
- 2\I e_a{}^\mu e_b{}^ie_c{}^\nu[A_\nu,A_\mu]\Phi_i
-2 e_a{}^\mu e_b{}^i e_c{}^\nu f_{jki} \Omega_\mu{}^j A_\nu\Phi^k\right)
\nonumber \\
=\ &f^{abc}\left(-e_a{}^\mu e_b{}^i e_c{}^\nu F_{\mu\nu} \Phi_i
- \I e_a{}^\mu e_b{}^i e_c{}^\nu [A_\nu,A_\mu] \Phi_i
- e_b{}^i f_{ac}{}^{d} e_d{}^\nu A_\nu \Phi_i
+ B_{ab}^i e_c^\nu A_\nu \Phi_i\right)
\nonumber \\
=\ &f^{abc}\left(-e_a{}^\mu e_b{}^ie_c{}^\nu F_{\mu\nu}\Phi_i
- \I e_a{}^\mu e_b{}^i e_c{}^\nu [A_\nu, A_\mu] \Phi_i\right) \, .
\end{align}
The symbol ``$\simeq$'' on the first equality above refers to the omission of
a total derivative term.
Substituting this into eq.(\ref{eq:CScomponent}),
we find
\begin{align}
S_{CS}=&- \frac{3(L+2\lambda) L^3}{4} \int d\Theta \;
\mbox{Tr}\Biggl[
f^{abc}\biggl(
e_a{}^\mu e_b{}^\nu e_c{}^\rho(F_{\mu\nu}A_\rho
+ \frac{\I}{3}[A_\mu,A_\nu]A_\rho)
+ 2e_a{}^\mu e_b{}^\nu e_c{}^iF_{\mu\nu}\Phi_i\biggr)
\nonumber \\
&\hspace{10.2em}
- \frac{2}{3}\I f^{ijk}[\Phi_i,\Phi_j]\Phi_k
- (2h^{\mathfrak{h}}_{ij}-h^{\mathfrak{g}}_{ij}) \Phi^i \Phi^j \Biggr].
\label{eq:CScomponent2}
\end{align}

Combining $S_{\rm kinetic}$ (\ref{eq:kineticcomponent})
and $S_{CS}$ (\ref{eq:CScomponent2}),
we have the total action:
\begin{equation}
\begin{split}
S=&
L^4 \int d\Theta \;
\mbox{Tr}\Biggl[
- \frac{1}{4}\left(F_{\mu\nu}F^{\mu\nu}
+ 2\nabla_\mu\Phi_i\ \nabla^\mu\Phi^i
- [\Phi_i,\Phi_j]^2\right)
\\
&\hskip5.1em
+ \frac{(3h^{\mathfrak{h}}_{ij}-2h^{\mathfrak{g}}_{ij}) + 3\frac{\lambda}{L}(2h^{\mathfrak{h}}_{ij}-h^{\mathfrak{g}}_{ij})}{2}\Phi^i \Phi^j
+ \I \frac{\lambda}{L} f^{ijk}\Phi_i[\Phi_j,\Phi_k]
\\
&\hskip5.1em
- f^{abc}\left(
\left(2+3\frac{\lambda}{L}\right)e_a{}^\mu e_b{}^\nu e_c{}^i F_{\mu\nu} \Phi_i
+ \frac{3\left(1+2\frac{\lambda}{L}\right)}{4}e_a{}^\mu e_b{}^\nu e_c{}^\rho
\left(F_{\mu\nu}A_\rho+\frac{\I}{3}[A_\mu,A_\nu]A_\rho\right)\right)
\Biggr] \, .
\end{split}
\label{total-action}
\end{equation}
This is an action for the gauge potential $A_{\mu}$
and $(|G|-|H|)$ massive scalar fields $\Phi_i$
in the adjoint representation of the gauge group $K$
living on a coset space $G/H$.
It includes the Yang-Mills action
and a Chern-Simons-like term.
The massive scalars $\Phi$ are
non-minimally coupled to the gauge field,
and has cubic and quartic self-interactions.
The action is invariant under both
the gauge group $K$ and the global symmetry $G$
as an isometry of the base space $G/H$.

Apart from an overall scaling $L^4$ of the total action
which corresponds to the Yang-Mills coupling $g_{YM} = 1/L^2$,
the only other coupling constant is the dimensionless quantity $\lambda/L$.
The mass squared of the scalar fields is
\begin{align}
M^2_{ij} \equiv
\frac{(2h^{\mathfrak{g}}_{ij} - 3h^{\mathfrak{h}}_{ij}) + 3\frac{\lambda}{L}(h^{\mathfrak{g}}_{ij} - 2h^{\mathfrak{h}}_{ij})}{2} \, ,
\end{align}
which does not only depend on the coupling $R^2 = L(L+3\lambda)/2$
of the quadratic term $S_1$ in the action (\ref{action-modified}).

While a generic action of this form admits 
many more independent coupling constants,
this action is special in being
that of a pure gauge theory
(with the Yang-Mills term and the Chern-Simons-like term)
dimensionally reduced to the coset space $G/H$,
apart from the mass term.

In the above,
we have assumed that the number of matrices 
in the model equals $|G|$.
To relax this assumption,
one can simply replace $G$ and $H$ by
$G' = G \times U(1)^n$ and $H' = H \times U(1)^n$,
so that $G'/H'$ is the same coset space as $G/H$.
The algebra $\hat{C}(K; G', H')$
is clearly different from $\hat{C}(K; G, H)$
as it has a larger number ($|G| + n$) of matrices.

We use $a, b, c$ as labels of the generators of $G$,
the symbols $\bar{a}, \bar{b}, \bar{c}$ as those of $U(1)^n$,
and $a', b', c'$ as those of $G'$.
We can simply take $Z_a$ given by eq.(\ref{eq:Zsolution2})
and $Z_{\bar{a}} = 0$ as the solution
to the equation of motion (\ref{Z-new-eq}) for $Z_{a'}$,
since the structure constant $f_{a'b'}{}^{c'}$ vanishes
when any of the indices take values in the $U(1)^n$ factor,
i.e.
\begin{align}
f_{\bar{a}b'}{}^{c'} = f_{a'\bar{b}}{}^{c'} = f_{a'b'}{}^{\bar{c}} = 0 \, .
\end{align}

We introduce the index $\bar{i}$ to label the additional scalar fields $\Phi_{\bar{i}}$
corresponding to the extra $U(1)^n$ factor added to both $G$ and $H$,
so that eq.(\ref{X-comp}) is now
\begin{align}
\hat{X}_{a'} = e_{a'}{}^{\mu} A_{\mu} + e_{a'}{}^i \Phi_i + e_{a'}{}^{\bar{i}} \Phi_{\bar{i}} \, .
\end{align}
The basis $e_{a}{}^{\mu}$, $e_{a}{}^{i}$ are defined as before,
and the new components are defined by
\begin{eqnarray}
e_{a}{}^{\bar{i}} =
e_{\bar{a}}{}^{\mu} = 
e_{\bar{a}}{}^{i} = 0 \, ,
\qquad
e_{\bar{a}}{}^{\bar{i}} = \delta_{\bar{a}}{}^{\bar{i}} \, .
\end{eqnarray}
It is then easy to see that,
when the algebra $\hat{C}(K; G, H)$ is replaced by $\hat{C}(K; G', H')$,
the total action is simply eq.(\ref{total-action})
supplemented by the addition terms
\begin{align}
S_{add.} &= L^4\int d\Theta \, \mbox{Tr}\left(
- \frac{1}{2} \nabla_{\mu}\Phi_{\bar{i}}\nabla^{\mu}\Phi_{\bar{i}}
+ \frac{1}{4} [\Phi_{\bar{i}}, \Phi_{\bar{j}}]^2+\frac12 [\Phi_i, \Phi_{\bar{i}}]^2
\right)
\end{align}
for $n$ new scalar fields $\Phi_{\bar{i}}$
in the adjoint representation
minimally coupled to the gauge fields.

\section{Conclusion and Discussion}
In this paper, we propose an infinite-dimensional algebra $\hat C(K; G, H)$ to describe a gauge theory with the gauge group K on the coset space $G/H$. The IKKT model with the extra mass term and the Chern-Simons term produces a gauge-invariant Lagrangian on the coset with the scalar fields associated with $H$.

This formulation allows us to discuss interesting base spaces
such as the de Sitter space and the anti-de Sitter space.
As the metric of the model needs to coincide 
with the invariant non-degenerate inner product on $\mathfrak{g}$,
the metric of the model needs to be
$\eta = \mbox{diag}(-1, 1, \cdots, 1)$,
as the original IKKT model,
for the de Sitter space $dS^d = O(d, 1)/O(d-1, 1)$.
For the anti-de Sitter space
$AdS_d = O(d-1, 2)/O(d-1, 1)$,
the metric of the model should be
$\eta = \mbox{diag}(-1, -1, 1, \cdots, 1)$.

While we limit the analysis to the IKKT-type model for simplicity,
one may use $\hat C(K; G, H)$ as a gauge group of
the gauge theories in arbitrary dimensions,
e.g. the BFSS matrix model \cite{Banks:1996vh}.

There are a few open questions to be explored.
\begin{itemize}

\item When we use $\hat C(K; G, H)$ for the higher dimensional gauge theory with a base manifold $M$, the coset space becomes a fiber over $M$. In the expansion of the gauge fields, there appears a connection on $M$, which describes the fiber bundle \cite{Ho:2009nk}. 
It will be interesting to extend our formulation to these more general cases.

\item What is the algebra associated with a generic manifold $M$? A natural subset of generators is the gauge algebra-valued functions on $M$. We also need an analogue of the differential operators as members of the algebra and their pair partners. The requirement of the Jacobi identity and the invariance of the inner product impose strict constraints. 
\footnote{
Here we are concerned with an ordinary gauge theory.
For a higher-spin gauge theory,
a generic curved spacetime in the IKKT matrix model \cite{IKKT}
has been formulated in Refs.\cite{Hanada:2005vr,Kawai:2007tn,Sakai:2019cmj}.
}

\item What is the representation theory for $\hat C$, and what is its role? For $\hat C(K;  U(1), id)$, the algebra is very well-known (Kac-Moody algebra) and has been well-studied. Recently, the quantum deformation of $\hat C(K, U(1)^{\otimes 2}, id)$ has been studied in the context of the topological strings. For other cases, even the simple case of $G/H=S^2$, there are almost no results about the representation theory.

\item
It is of interest to study the extension of the algebra $\hat{C}$ to a matrix algebra.
This can be done in two directions.
One direction is to let the field variables to take values in the universal enveloping algebra of $\hat{C}$,
so that higher-spin fields are included through higher-derivative terms,
along the lines of Refs.\cite{Hanada:2005vr,Kawai:2007tn,Sakai:2019cmj}.
The other direction is to construct a non-commutative coset space
whose algebra of functions is the matrix algebra for a finite $N$,
and then recover the commutative space in the large $N$ limit.

\end{itemize}

We leave these questions for future study.

\section*{Acknowledgement}

KH is supported in part by JSPS fellowship.
PMH is supported in part by the National Science Council, Taiwan, R.O.C.
and by National Taiwan University.
YM is partially supported by Grant-in-Aid (KAKENHI \#18K03610 and \#18H01141)
from MEXT Japan.
AW is partially supported by JSPS fellowship.

\appendix

\section{Check of Jacobi identity and invariance of inner product}\label{a:Jacobi}
In this section, 
we check the Jacobi identity and the invariance of the inner product. 
We first consider the defining relation (\ref{eq:extension}) for the central extension.
From eq.(\ref{eq:extension}), we have 
\ba
\left[[\cT^{\alpha}[\lambda_1],\cT^{\beta}[\lambda_2]],\cT^{\delta}[\lambda_3]\right]=-{F^{\alpha\beta}}_{\gamma}{F^{\gamma\delta}}_{\sigma}\cT^{\sigma}[\lambda_1\lambda_2\lambda_3]
-{\rm i}F^{\alpha\beta\delta}v_a\int d\Theta \, \lambda_1\lambda_2\hat{u}^a\lambda_3 \, .
\ea
We do not need to consider the first term because
the Jacobi identity is satisfied for any Lie group $K$. 
For the second term, 
the Jacobi identity can be shown from the cyclic symmetry of $F^{\alpha\beta\delta}$
and the following relation 
\ba
\int  d\Theta \, (\lambda_1\lambda_2\hat{u}^a\lambda_3+\lambda_2\lambda_3\hat{u}^a\lambda_1+\lambda_3\lambda_1\hat{u}^a\lambda_2)=\int  d\Theta \  \hat{u}^a(\lambda_1\lambda_2\lambda_3)=0 \, .
\ea
Thus, the central extension is consistent with the Jacobi identity.

Next, we need to check that the action of $u^a$ is consistent.
The action of $u^a$ on $T^{\alpha}[\lambda]$ and $v_a$ is obvious,
so we only have to check its consistency with eq.(\ref{eq:extension}).
This can be checked as follows:
\begin{equation}
\begin{split}
&\left[[\hat{u}^a, \cT^{\alpha}[\lambda_1]], \cT^{\beta}[\lambda_2]\right]+\left[\cT^{\alpha}[\lambda_1],[\hat{u}^a,\cT^{\beta}[\lambda_2]]\right]\\
=\ &{\rm i}{F^{\alpha\beta}}_{\gamma}\cT^{\gamma}[(\hat{u}^a\lambda_1)\lambda_2+\lambda_1\hat{u}^a\lambda_2]-G^{\alpha\beta}v_b\int d\Theta \, (\hat{u}^a\lambda_1)(\hat{u}^b\lambda_2)-G^{\alpha\beta}v_b\int d\Theta \,  \lambda_1\hat{u}^b\hat{u}^a\lambda_2\\
=\ &{\rm i}{F^{\alpha\beta}}_{\gamma}\cT^{\gamma}[\hat{u}^a(\lambda_1\lambda_2)]+G^{\alpha\beta}v_b\int d\Theta \, \lambda_1\hat{u}^a\hat{u}^b\lambda_2-G^{\alpha\beta}v_b\int d\Theta \, \lambda_1\hat{u}^b\hat{u}^a\lambda_2\\
=\ &{\rm i}{F^{\alpha\beta}}_{\gamma}\cT^{\gamma}[\hat{u}^a(\lambda_1\lambda_2)]+{\rm i}{f^{ab}}_cG^{\alpha\beta}v_b\int d\Theta \, \lambda_1\hat{u}^c\lambda_2\\
=\ &{\rm i}{F^{\alpha\beta}}_{\gamma}\cT^{\gamma}[\hat{u}^a(\lambda_1\lambda_2)]-G^{\alpha\beta}[u^a,v_c]\int d\Theta \, \lambda_1\hat{u}^c\lambda_2\\
=\ &\left[\hat{u}^a, [\cT^{\alpha}[\lambda_1], \cT^{\beta}[\lambda_2]]\right] \, .
\end{split}
\end{equation}
We have thus completed the consistency check for the algebra $\hat C(K; G, H)$.  

Next, we should check the invariance of the inner product (\ref{eq:Killing1},\ref{eq:Killing2}).
Using eq.(\ref{Xcomp}),
we have
\begin{equation}
\begin{split}
&<[\cX_I,\cX_J],\cX_K>\\
&= \ {\rm i} X_{IA\Xi_1}X_{JB\Xi_2} F^{ABC} X_{KC\Xi_3}
\int d\Theta \, \lambda_{\Xi_1} \lambda_{\Xi_2} \lambda_{\Xi_3}
\\
&- X_{IA\Xi_1}Z_{Ja}G^{AB}X_{KB\Xi_2}
\int d\Theta \, (\hat{u}^a\lambda_{\Xi_1}) \lambda_{\Xi_2}
+X_{JA\Xi_1}Z_{Ia}G^{AB}X_{KB\Xi_2}
\int d\Theta \, (\hat{u}^a\lambda_{\Xi_1})\lambda_{\Xi_2}
\\
&+{\rm i}Z_{Ia}Z_{Jb}{f^{ab}}_cY_K^c
-X_{I\alpha\Xi_1}X_{J\beta\Xi_2}G^{\alpha\beta}Z_{Ka}\int d\Theta \, \lambda_{\Xi_1}\hat{u}^a\lambda_{\Xi_2}
+{\rm i}{f^{bc}}_aY_I^aZ_{Jb}Z_{Kc}-{\rm i}{f^{bc}}_aY^a_JZ_{Ib}Z_{Kc} \, .
\end{split}
\end{equation}
One can see that this expression is anti-symmetric for the indices $J,K$. We note that the sum of the first term in the second line and the term in the fourth line form an anti-symmetric term as well as that of the term in the third line and the second term in the last line. The anti-symmetry for $J,K$ and the symmetric nature of the inner product lead to invariance under the action of the algebra,
\ba
<[\cX_I,\cX_J],\cX_K>+<\cX_J,[\cX_I,\cX_K]>=0 \, .
\ea

\hide{
\section{Differential geometry for general coset}\label{a:general}

\subsection{Connection}

The connection one-form $\Gamma^\alpha_\beta$ is defined by
\ba
dV^\alpha +\Gamma^\alpha_\beta\wedge V^\beta=0\,.
\ea
In order to have the torsion free connection, we need to impose the antisymmetry for the connection,
\ba
\Gamma^\alpha_\beta \delta^{\beta\gamma}=-\Gamma^\beta_\gamma \delta^{\gamma\alpha}\,.
\ea
The compatibility with the Maurer-Cartan equation gives,
\ba
\Gamma^\alpha_\beta =\frac\I2\left(
-{f_{\beta\gamma}}^\alpha+\delta_{\beta\epsilon} {f_{\delta\gamma}}^\epsilon \delta^{\alpha\delta}
+\delta_{\gamma\epsilon} {f_{\delta\beta}}^\epsilon \delta^{\alpha\delta}\right)V^\gamma-\I {f_{\beta i}}^\alpha \Omega^i\,.
\ea
\subsection{Curvature}
See \cite{Camporesi:1990wm}.
}

\section{Coset representation of $S^2$}\label{cosetS2}

One may take the spherical coordinates $\{\theta^{\mu}\} = \{\theta,\varphi\}$
and define
\ba
\int d\Theta &=& \int_0^{\pi} \sin\theta d\theta\int_0^{2\pi}d\varphi \, ,\\
\hat u_1 &=&  {\rm i}(\sin\varphi \,\partial_\theta+\cos\varphi\cot\theta\,\partial_\varphi) \, , \label{u1}\\
\hat u_2 &=& -{\rm i}(\cos\varphi \,\partial_\theta-\sin\varphi\cot\theta\,\partial_\varphi) \, ,
\label{u2}\\
\hat u_3&=& -\I\,\partial_\varphi \, ,
\label{u3}
\\
{f_{ab}}^c&=&\epsilon_{abc},\quad g^{ab}=\delta^{ab}\,.
\ea
One may identify the basis of the orthonormal functions on $S^2$ by the spherical harmonics $Y_{\ell m}$\,.

We describe the explicit coset of $SU(2)/U(1)$. The generator of $SU(2)$ is described by $T_a =\tau_a/2$
($\tau_a$ is the Pauli matrices) and the subgroup $K=U(1)$ is generated by $T_3$.  The element of the coset is described by $\sigma(\theta,\varphi) H$ with $\sigma(\theta,\varphi)\in SU(2)$ and $h=K=U(1)$. We introduce coordinates $\theta,\varphi$ of the coset by
\begin{eqnarray}
&&\sigma(\theta,\varphi)=\exp\left(
\I \sum_{\mu=1,2} t_\mu T_\mu
\right)=\left(
\begin{array}{cc}
\cos(\theta/2) & -e^{-\I \varphi}\sin(\theta/2)\\
e^{\I \varphi}\sin(\theta/2) & \cos(\theta/2)
\end{array}
\right),\\
&&\theta=\sqrt{t_1^2+t_2^2},\quad 
\varphi=-\arctan(t_1/t_2)\,.
\end{eqnarray}
We note that it satisfies
\begin{eqnarray}
\sigma(\theta,\varphi)^{-1}=\sigma(-\theta,\varphi)
\end{eqnarray}

The vierbein and the $H$-connection are given by
\ba
\sigma^{-1}d\sigma=\I V^{\alpha}_\mu T_\alpha+ \I \Omega_\mu T_3 \, ,
\ea
with
\ba
V_\theta^1=\sin\varphi,\quad
V_\theta^2=-\cos\varphi,\quad
V_\varphi^1=\cos\varphi\sin\theta,\quad
V_\varphi^2=\sin\varphi\sin\theta,\quad
\Omega_\theta =0,\quad
\Omega_\varphi=1-\cos\theta\,.
\ea
The metric tensor is
\ba
ds^2 =\delta_{ab}V^a_\mu V^b_\nu d\theta^\mu d\theta^\nu=(d\theta)^2+\sin^2\theta(d\varphi)^2\,.
\ea

The adjoint matrix is given by
\begin{eqnarray}
\sigma^{-1} T_a \sigma ={D_{a}}^b(\sigma) T_b
\end{eqnarray}
with
\begin{eqnarray}
D(\sigma)=\left(
\begin{array}{ccc}
\cos ^2\left(\frac{\theta }{2}\right)-\cos (2 \varphi ) \sin ^2\left(\frac{\theta }{2}\right) & -\sin
^2\left(\frac{\theta }{2}\right) \sin (2 \varphi ) & \cos (\varphi ) \sin (\theta ) \\
-\sin ^2\left(\frac{\theta }{2}\right) \sin (2 \varphi ) & \cos ^2\left(\frac{\theta }{2}\right)+\cos (2 \varphi ) \sin
^2\left(\frac{\theta }{2}\right) & \sin (\theta ) \sin (\varphi ) \\
-\cos (\varphi ) \sin (\theta ) & -\sin (\theta ) \sin (\varphi ) & \cos (\theta ) \\
\end{array}
\right) \, .
\end{eqnarray}
We note that the third line is identical to the $\vec e^r$, ${D_a}^3=(\vec e^{r})_a$.
One may obtain the other polar basis by contraction with the inverse vierbein,
\begin{eqnarray}
{e_a}^\mu=(\vec e^\theta, \sin^{-1}(\theta)\vec e^\varphi)={D_a}^\alpha {V_\alpha}^\mu =
\left(
\begin{array}{cc}
\sin (\varphi ) & \cot (\theta ) \cos (\varphi ) \\
-\cos (\varphi ) & \cot (\theta ) \sin (\varphi ) \\
0 & -1 \\
\end{array}
\right)
\end{eqnarray}
The orthogonality of the polar basis comes from that of the adjoint matrix ${D_a}^b(\sigma){D_c}^d(\sigma)\delta^{ac}=\delta^{bd}$. In particular,
\begin{eqnarray}
\delta^{ab} {e_a}^\mu {e_b}^\nu=g^{\mu\nu}, \quad
\delta^{ab}{e_a}^\mu {e_b}^i=0,\quad \delta^{ab}{e_a}^i {e_b}^j=\delta^{ij} \, .
\end{eqnarray}
For $SU(2)/U(1)$ case, 
the indices $i, j$ are limited to be equal to $3$ and ${e_a}^3={\vec e^r}_a={D_a}^3$.

The isometry of the coset may be realized by the left action of $T_a$. One may derive the following identity,
\ba
T_a \sigma(\theta,\varphi) =- \hat u_a \sigma(\theta,\varphi)+ 
\Lambda_a  \sigma(\theta,\varphi) T_3
\ea
where $\hat u_a={u_a}^\mu\partial_\mu =\I {e_a}^\mu\partial_\mu$ are given in eqs.(\ref{u1},\ref{u2},\ref{u3}) and
\ba
\Lambda_1=\cos\varphi\tan(\theta/2)\, , \quad
\Lambda_2=\sin\varphi\tan(\theta/2)\, ,\quad
\Lambda_3=1 \, .
\ea

\end{document}